1 February 2014

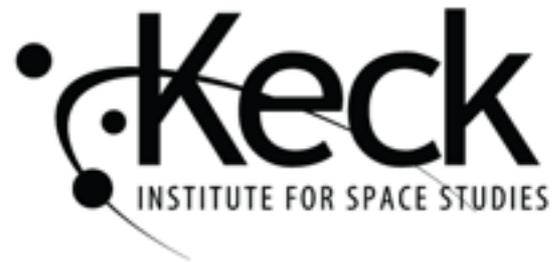

# *AIRSHIPS:*
# A New Horizon for Science

Study report prepared for **The Keck Institute for Space Studies**

*www.kiss.caltech.edu/study/airship/*

Opening Workshop:     April 30-May 3, 2013
Closing Workshop:     September 25-27, 2013
Summit on Tethers:    November 18, 2013

### Study Co-leads:
Sarah Miller (UCI/Caltech), Robert Fesen (Dartmouth), Lynne Hillenbrand (Caltech), Jason Rhodes (JPL)

### Study Participants:
Gil Baird (ILC Dover), Geoffrey Blake (Caltech), Jeff Booth (JPL), David E. Carlile (Lockheed Martin ADP), Riley Duren (JPL), Frederick G. Edworthy (Aeros), Brent Freeze (Sorlox Corporation), Randall R. Friedl (JPL), Paul F. Goldsmith (JPL), Jeffery L. Hall (JPL),  Scott E. Hoffman (Northrop Grumman), Scott E. Hovarter (Lockheed Martin Space Systems), Rebecca M. Jensen-Clem (Caltech), Ross M. Jones (JPL), Jens Kauffmann (Caltech), Alina Kiessling (JPL), Oliver G. King (Caltech), Nick Konidaris (Caltech), Timothy L. Lachenmeier (Near Space Corporation), Steven D. Lord (Caltech/IPAC), Jessica Neu (JPL), Gregory R. Quetin (University of Washington), Alan Ram (Northrop Grumman), Stanley Sander (JPL), Marc Simard (JPL), Mike Smith (Raven Aerostar), Steve Smith (Southwest Research Institute), Sara Smoot (Stanford), Sara Susca (JPL), Abigail Swann (University of Washington), Eliot F. Young (Southwest Research Institute), Thomas Zambrano (AeroVironment, Inc.)

# Table of Contents

# Executive Summary

The *Airships: A New Horizon for Science* study at the Keck Institute for Space Studies investigated the potential of a variety of airships currently operable or under development to serve as observatories and science instrumentation platforms for a range of space, atmospheric and Earth science. The participants represent a diverse cross-section of the aerospace sector, NASA, and academia. They are leaders in their respective fields who have built or are building high altitude airships, or are Earth, atmospheric, planetary, or astrophysics scientists interested in exploiting airship platforms.

Over the last two decades, there has been wide interest in developing a high altitude, stratospheric lighter-than-air (LTA) airship that could maneuver and remain in a desired geographic position (i.e., "station-keeping") for weeks, months or even years. Such a stratospheric airship would offer the military surveillance capabilities over large areas. This platform would also provide telecommunication companies a means of providing commercial communication and data services to consumers in remote areas.  While stratospheric airships remain a promise rather than a reality today, seeing through the final stages of development of such vehicles operating in the relatively light winds present in the lower stratosphere at altitudes around 65 kft (20 km), would enable unique data collection opportunities for Earth and atmospheric scientists. They would be a game-changer for space scientists since their costs as a platform would be substantially lower than satellite missions.

The original goals of the study were to:

1) **Inform** scientists of the capabilities of airship vehicles as instrumental platforms, as well as discuss how this technology could be expanded and improved to better accommodate science instrumentation requirements.

2) **Identify** science observational/experimental projects that are uniquely addressed by airship vehicles, as well as science which can be supported by airships at a significantly lower cost than other platforms (i.e., satellites).

3) **Construct** science concepts for viable airship platforms.

Our study found considerable scientific value in both low altitude (< 40 kft) and high altitude (> 60 kft) airships across a wide spectrum of space, atmospheric, and Earth science programs. An airship provides persistent, high-resolution measurements that fill an observational scale gap in Earth and atmospheric science between "anecdotal" ground-based or aircraft measurements and coarse-resolution satellite measurements.  In addition, in situ and remote sensing views of our dynamic and evolving atmosphere, Earth ecosystems, coastal processes, atmospheric plume chemistry, extreme weather, upper troposphere and lower stratosphere processes like convection and exchange across the tropopause could all be made possible using airships.  Airships also open up the parameter space of long-duration, high spatio-temporal resolution observations of "Urban Dome" air quality associated with large cities. For instance in astrophysics, a 1-2 meter optical telescope placed at about 65 kft with state-of-the-art pointing stability would have superior resolving power to any optical ground-based telescope, providing exceptional image quality night after night above the weather.

While free-flying stratospheric balloons enable a wide range of observations, they do not satisfy the station-keeping needs of some applications, nor the long duration, global access needs of others. Over the course of the study period, we identified stratospheric tethered aerostats as a viable alternative to airships where station-keeping was valued over maneuverability.

By opening up the sky and Earth's stratospheric horizon in affordable ways with long-term flexibility, airships allow us to push technology and science forward in a project-rich environment that complements existing space observatories as well as aircraft and high-altitude balloon missions. Science, rather than war, could be the ultimate motivator to push industry toward final development of stratospheric airships, which will provide a unique platform for monitoring our most precious resource, the Earth, and for seeking out the new cosmic horizons toward the edge of our observable universe.

## Summary of Recommendations

The study concluded with three follow-on recommendations:

I. **Build a roadmap to stratospheric airship observatories:** Establish a "Challenge/Prize" for the development of a maneuverable, station-keeping, stratospheric airship, which can stay aloft at an altitude above 65 kft (20 km) for at least a full diurnal cycle (>20 hours) while carrying a science payload of at least 20 kg in mass.

II. **Take advantage of existing low and mid altitude airships:** Develop a consortium led by atmospheric and Earth science users to make immediate use of existing low altitude airships for a wide variety of science programs.

III. **Pursue technical development of stratospheric tethered platforms:** Construct and fly one or more LTA stratospheric vehicles tethered to the ground, a sea vessel, or even a secondary lower altitude aerial vehicle, to test their use as science platforms.

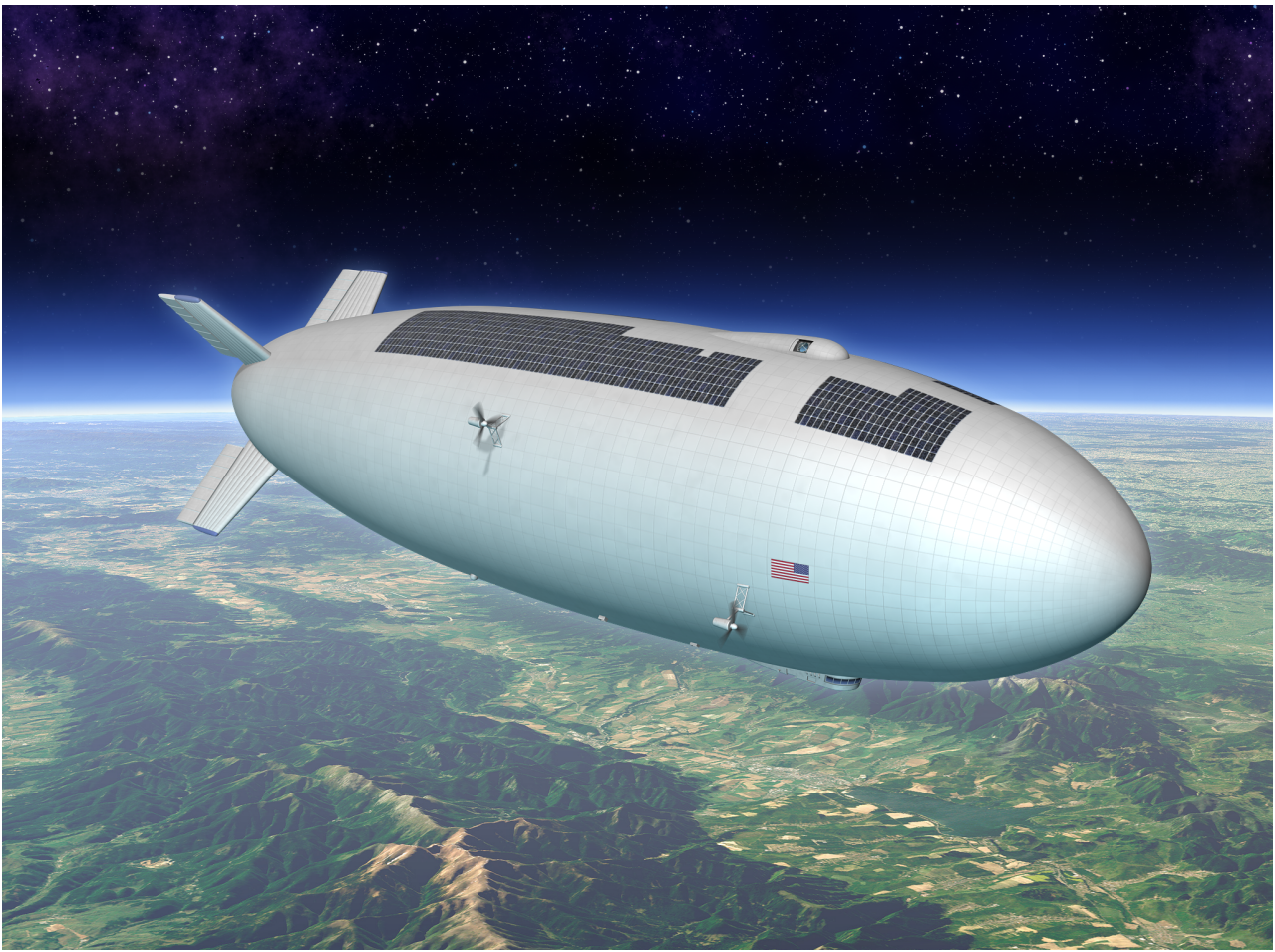

*Illustration of an airship observatory concept, including a world-class telescope mounted on the top of the airship and a suite of Earth and atmospheric instruments mounted on the bottom. Mike Hughes (Eagre Interactive) / Keck Institute for Space Studies*

# Chapter 1: Airships as a Platform for Science

## Challenges for Earth Science

Future Earth science efforts must untangle profoundly complex, large-scale natural and human interactions in pursuit of understanding and predicting environmental change impacts at local and regional scales. The greatest impediment to these efforts is in successfully connecting phenomena across a wide range of temporal and geographical scales.  A platform enabling **persistent, high resolution, local-to-regional scale observations** would fill a critical niche within Earth observing platforms.

The potential for such a platform to advance Earth science can be understood within the context of existing Earth observing platforms. Iconic, space-based images of the Earth, such as *Earth Rise*, *Pale Blue Dot*, and *Blue and Black Marbles* have not only transformed human perceptions of our Planet but have revealed a complex "system" of interactions among the atmosphere, hydrosphere, biosphere, and lithosphere, as well as human activities.

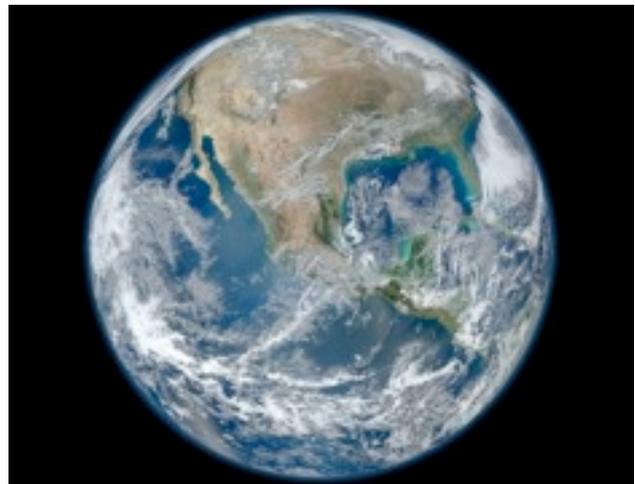

*Image credit: NASA*

As highlighted in the latest Decadal Survey report regarding Earth sciences from the National Research Council (NRC, 2007 - Earth), the unique capabilities of space-based observations are proving to be essential in discovering and understanding large-scale Earth system issues such as:

- The extent and chemistry of the ozone hole
- The transport of air pollution between countries and continents
- The rates of glacial and sea ice retreat
- Changes in land cover and use due to both human and natural causes
- Changing weather patterns due to pollution and land conversion
- The complex interactions between earthquake faults
- The global-scale effects of El Niño and La Niña on weather and the ocean's state and productivity
- The development and tracking of hurricanes, typhoons, and other severe storms
- Assessing damage from natural disasters and targeting relief.

Satellite observations, in conjunction with key ground and airborne data, are showing that global-scale change is occurring in the Earth system at an unprecedented rate.  In particular such observations reveal that:

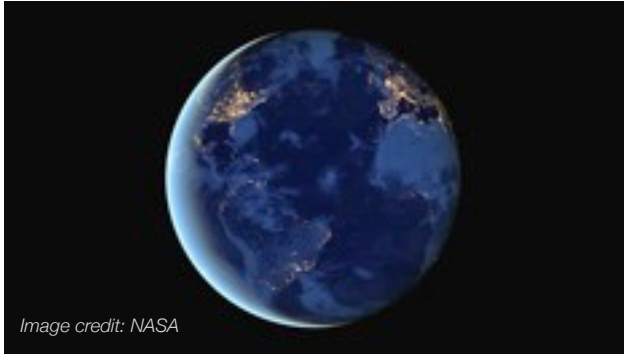
*Image credit: NASA*

- Human impacts on the Earth system are readily observable from space
- Carbon dioxide levels have risen to historic levels
- Sea level is rising at an accelerating pace
- Snow cover, sea ice, and glaciers are shrinking at unprecedented rates
- Arctic ecosystems are undergoing rapid change

While satellite measurements provide a global perspective on these rapid changes in the Earth-ocean-atmosphere system and ground and airborne data provide local or short-term measurements at fine spatiotemporal scales, we lack long-term, persistent measurements that can provide a critical link between small scale processes and regional and global long-term change.

## Challenges for Space Science

Many frontier areas of space sciences, planetary physics, astrophysics and cosmology rely on being above at least 95% of the atmosphere for key science acquisition. The Hubble Space Telescope has been one of the most successful space science enterprises in human history. With its invariably high over-subscription rate across its extended life-time, the astronomical community will suffer a great loss when this UV-optical observatory is finally extinguished. Hubble has enabled Noble-prize winning science regarding the expansion rate of the universe and dark energy, the discovery of the most distant galaxies, and the deepest probes of matter, including dark matter, in the universe. While the astronomical community eagerly anticipates the 6.5-m James Webb Space Telescope (2018 launch) for unprecedentedly crisp and sensitive imaging and spectroscopy in the infrared and red optical light, currently no open space-based observatory-class facility capable of acquiring wavelengths short-ward of the red part of the

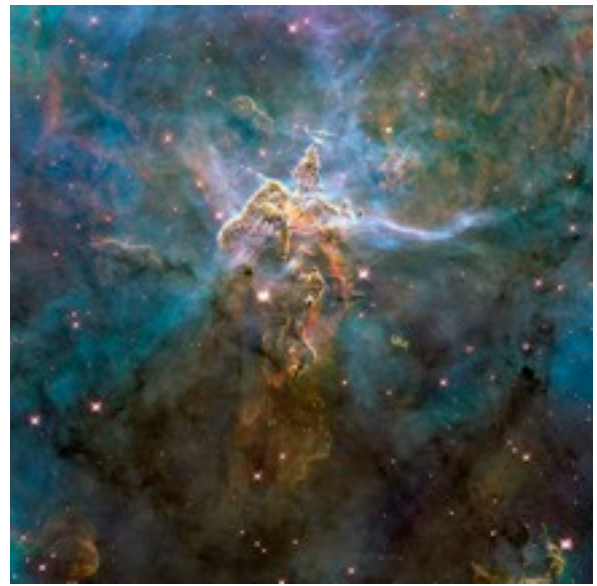
*Image Credit: NASA, ESA, and M. Livio and the Hubble 20th Anniversary Team (STScI)*

electromagnetic spectrum is planned to take Hubble's place. The European Space Agency will launch the highly anticipated Euclid space telescope (1.2-m) in 2020, covering near-infrared and optical wavelengths down to 550 nm, however the program of this Medium Class mission is set over 6 years and thus is not an open facility for the community like Hubble.

With a similarly high over-subscription rate to Hubble, the Atacama Large Millimeter Array (ALMA) telescope is revolutionizing far-infrared and millimeter technology. However, large swaths of the spectrum to which ALMA could be sensitive are blocked out by the Earth's atmosphere. Even the Stratospheric Observatory for Infrared Astronomy (SOFIA) observatory cannot access spectral regions absorbed by the atmosphere above its operating altitude of ~40 kft, leaving a significant fraction of the Terahertz sky unexplored. It is likely that precious clues to understanding the births of stars, exosolar planetary systems, exoplanet atmospheres, and signatures of life beyond Earth await us in these uncharted regions of the spectrum. Having a platform just 20 kft higher than SOFIA would make a remarkable difference to what space scientists can observe.

Additionally, astrophysicists require more flexible technology development in space-like conditions without space-like difficulty. For instance the new detector technology needed for probing the epoch of inflation and the extremely high-redshift early universe would be greatly aided by a flexible, accessible, near-space platform, saving millions in development.

To answer the big, outstanding questions faced by space scientists, their communities increasingly rely on massive projects and collaborations, which ultimately limit programmatic flexibility as well as the holistic training of future research leaders. Clearly underlined in the National Research Council Astronomy and Astrophysics Decadal Survey (NRC Astro2010) is the need for a hierarchy of project sizes, both to train our next generation as well as to allow for more immediate and flexible investigations.

## A Compelling Opportunity for Scientists

Interest in the development and use of long duration, lighter-than-air (LTA) vehicles for science applications has grown considerably in recent years, especially those with the capability to maintain station over a desired geographic position ("station-keeping"). Conventional airships ("blimps") generally have a streamlined structure and gain their altitude through buoyancy rather than lift, with enclosed lightweight gas such as hydrogen or helium. Historically, airships have been flown at relatively low altitudes (< 10 kft) but have the distinct advantage over aircraft of being able to stay aloft and/or hover a long time without refueling and do so at a relatively low cost of energy consumption (see the recent historical review of airships by Liao and Pasternak 2009).

Ever since the development of high altitude balloons in the 1950's there has been a strong desire to develop higher altitude airships with the idea of a station keeping LTA vehicle that could operate for many days or even months. Such a vehicle could be powered by photovoltaic cells and operate in the relatively light winds in the lower stratosphere at altitudes around 65 kft.

There has been and currently is wide interest in such a high altitude airship. Wireless communications using high altitude LTA platforms for broadcast services have long been considered by telecommunication companies as a way to expand commercial communication and data services to consumers. These platforms could combine some of the best features of satellite and fixed wireless services such as short transmission delay times, small propagation loss, and relatively large service areas (Tozer and Grace 2001, Grace et al 2005). Programs such as the European HELINET and HAPCOS projects[1], the Japanese Stratospheric Airship Platform Study (Eguchi et al 1998), and the recent Google internet balloon project ("Project Loon")[2] are among some of the most recent efforts to use high-altitude (> 60 kft altitude) balloons for telecommunication purposes.

High altitude, station-keeping LTA platforms also have several obvious military applications. The United States Department of Defense (US DoD) alone has spent more than $500 million dollars over the last decade developing the necessary technology. Across all military services, there is an ever-increasing demand for real-time communications and "over-the-horizon" surveillance capabilities. LTA vehicles that could operate at stratospheric altitude could offer large surveillance areas and good

---

[1] www.hapcos.org

[2] www.google.com/loon

air defense survivability factors. For example, an airship at an altitude of 70 kft would have a line-of-sight regional coverage some 650 miles in diameter, meaning that just one such vehicle could, for example, survey nearly all of Afghanistan.

Furthermore, the potential for longer on-station times of airships compared to high altitude aircraft (like the U2 and the Global Hawk) along with a low probability of communication intercept due to stable, direct line-of-sight communications are major advantages of airships. In a way, an airship could function as a surrogate satellite but offer much shorter transmission distances and ranges and thus higher resolution for connectivity of ground transmitters and receivers. The most aggressive, well-funded series of efforts to develop a high-altitude airship have been pursued by the US DoD. A comprehensive review of these efforts and programs can be found in the October 2012 US Government Accountability Office Report GAO-13-81 and is summarized in this report in Chapter 2.

In this chapter, we introduce the tremendous scientific potential of airships as scientific platforms, including the development of a robust, high-altitude, lighter-than-air science platform that could maneuver and station-keep for many weeks to several months, and how airships fit into the existing suite of science platforms available today.

## Unique Airship Capabilities for Earth and Atmospheric Science

Earth and Atmospheric Science observations are currently carried out using ground measurements, static towers, ships, balloons, aircraft and satellites. These mature platforms allow for the observation of phenomena such as chemical emissions, weather dynamics and land cover change over many time scales and across diverse regions of the Earth. Each platform requires a trade-off between coverage and persistence. For example, an aircraft can map a region of aerosols or the biological emissions of a forest, but it only revisits every air parcel by circling back on a specific area. The process of circling back on an area sets a lower limit for the revisit time of that particular air parcel. A static tower can measure an area continuously but cannot follow a moving parcel of air and is limited by reachable and appropriate areas for placement. A balloon-mounted instrument might follow along with a parcel of air but the requirements for the launch location of the balloon and the control of its buoyancy will limit the geographical areas of study and persistence with the phenomenon of interest.

**Airships as scientific platforms allow new access to a set of phenomena not currently accessible by existing platforms.** In general terms this set of Earth and atmospheric sciences phenomena are characterized by the combination of difficult ground access or important flow-dependent evolution and the desire to observe with persistence on timescales of a day to a week. Some example phenomena that fit this description are urban, biological or geological atmospheric or ocean emission plumes (e.g. wild fires), remote ice dynamics and remote forest dynamics. The capabilities of an airship would allow for a relatively rapid response to the location of a dynamic event such as a wild fire and then drift with the plume while continuously measuring the rapid chemical reactions occurring in the plume. Such measurements are critical to understanding the transport of pollutants as well as the natural reactive processes occurring in the atmosphere. Alternatively a long endurance airship could enable persistent observations of regions around the Earth that have never before been explored in this manner, from the arctic sea ice to the tropical Amazon forest where over-land travel is prohibitively difficult. Both polar areas and the tropical rain forests are changing rapidly under climate change, and many processes critical for our understanding of their current and future health modulate over a daily cycle observable only in one location by towers established and operated at great cost. Some locations of interest in these areas are not presently accessible in any way. An airship has the potential for week-long endurance, mobility at multiple altitudes and the capability to hold position. Airship capabilities at low, medium and high altitude all have applications to open up parts of the world that have never been explored before in scientific detail.

Airships can fill a current gap between "anecdotal" (occasional or sporadic) ground-based observations and the lack of persistence and resolution from aircraft and satellite measurements, respectively. The capabilities of airships allow unique study of land, sea and air providing access to new regions of the planet. At extremely low altitudes for a fixed-wing aircraft, an airship can function as a mobile in-situ laboratory while at altitudes from 20–40 kft or > 60 kft, it can be a platform for co-location of an observatory of remote sensing instruments or in-situ measurements of intercontinental transport of pollution, exchange of water vapor and ozone-related gases across the tropopause, climate feedbacks on stratospheric dynamics, and/or springtime polar ozone loss.

## Advantages of Space Science from a High-Altitude Airship Platform

Cosmology, astronomy and astrophysics, planetary and other space sciences could greatly benefit from a persistent observing platform possessing space-like conditions, with capabilities that the current suite of NASA platforms does not offer. Astronomical space satellites are expensive, typically costing more than hundreds of millions of dollars, and often require extensive timelines to build. NASA's large high altitude balloons flown in or near the polar regions offer a less expensive and more rapid alternative but cannot take us everywhere scientists need to be, nor for as long as they wish to be there.

A stratospheric astronomy airship could provide space-like observing conditions across a broad range of wavelength regimes and be launched quickly, affordably, and repeatedly. Even a modest telescope at stratospheric altitudes would provide image quality that could compete with space-based telescopes (see Von Appen-Schnur and Luks 1998 describing the advantages of a high altitude observatory).

Large gains in atmospheric transmission, especially in the millimeter and sub-millimeter (THz) spectral region over ground-based observations, and even the SOFIA aircraft, are possible from altitudes of 60 kft or higher. Above 45 kft many advantages appear. This is the start of the tropopause, where liquid water freezes out of the atmosphere. The freezing-out of water allows far-infrared radiation to easily reach 60 kft, opening up the view of the "cool molecular Universe" photometrically and spectroscopically from 30-600 mm. Likewise, soft X-rays can penetrate downward to ~60 kft.

Airships will also permit long duration observing timeframes yielding high sensitivity via deep integrations and excellent point spread functions ("seeing"). In certain wavelength regimes, an airship-based telescope (compared to a ground-based observatory) would also enjoy nearly continuous "dark time" observing conditions due to greatly decreased sky brightness and scattering by moonlight. Airships could also provide rapid response for time-critical astronomical observations of a variety of astronomical events such as newly discovered comets, supernovae, gamma ray bursts, and other unpredictable transits.

## A True Platform Capability Gap Across Science

Airships represent an exciting complement and alternative to expensive geosynchronous earth orbiting (GEO) satellites or constellations of low earth orbit (LEO) satellites. A stable platform positioned in the lower or middle stratosphere (60-90 kft) would provide a space-like observation outpost far more accessible and less expensive than GEO or LEO platforms. Given an increasing number of well-motivated scientific satellite missions in the last three decades, there are strong drivers for the use of relatively inexpensive LTA vehicles for a wide range of Earth and space applications.

In particular for Earth science, the capabilities of stratospheric LTA platforms could be complementary to that of spacecraft (Smith and Rainwater 2003). While LEO satellites have proven to be extremely effective at capturing large-scale context, they do not provide persistent observations of specific localities or regions owing to their rapid orbital traverses. As a complement to LEO satellites, GEO satellites obtain continuous observations of specific regions but typically at the expense of degraded spatial resolution.  Neither GEO nor sun synchronous LEO satellites can capture diurnal behavior of targeted phenomena.  Also, given their higher cost and complexity, relatively few satellites are launched per year.  The low replenishment rate of NASA Earth satellites has been particularly acute over the past decade, with the present set of environmental satellites operating well beyond their design life, placing the system as a whole in danger of collapse (NRC, 2007 - Earth). These satellite systems cost on the order of 1 billion USD (10-100 times the cost of airships) and typically conduct specifically designed experiments on non-reusable platforms.

Recognizing the need for critical and affordable observations that span the range of Earth processes, the National Global Change Research Plan (USGCRP, 2012) seeks to "sustain and strengthen the capacity to observe long-term changes in the global Earth system and integrate observations to improve fundamental understanding of the complex causes and consequences of global change".  As part of that capacity, the NRC finds that alternative platforms, such as balloons and aerial vehicles, offer flexibility and may be employed, in some cases, to lower the cost, relative to satellites, of meeting science objectives (NRC, 2012).

Figure 1.1 compares the observational attributes of airships to other major platforms. In contrast to satellites, suborbital platforms obtain higher spatial resolution, capture diurnal behavior and, relative to LEO, provide more persistent local and regional observations.  However, these advantages typically come at the expense of large-scale spatial coverage and observational duration.

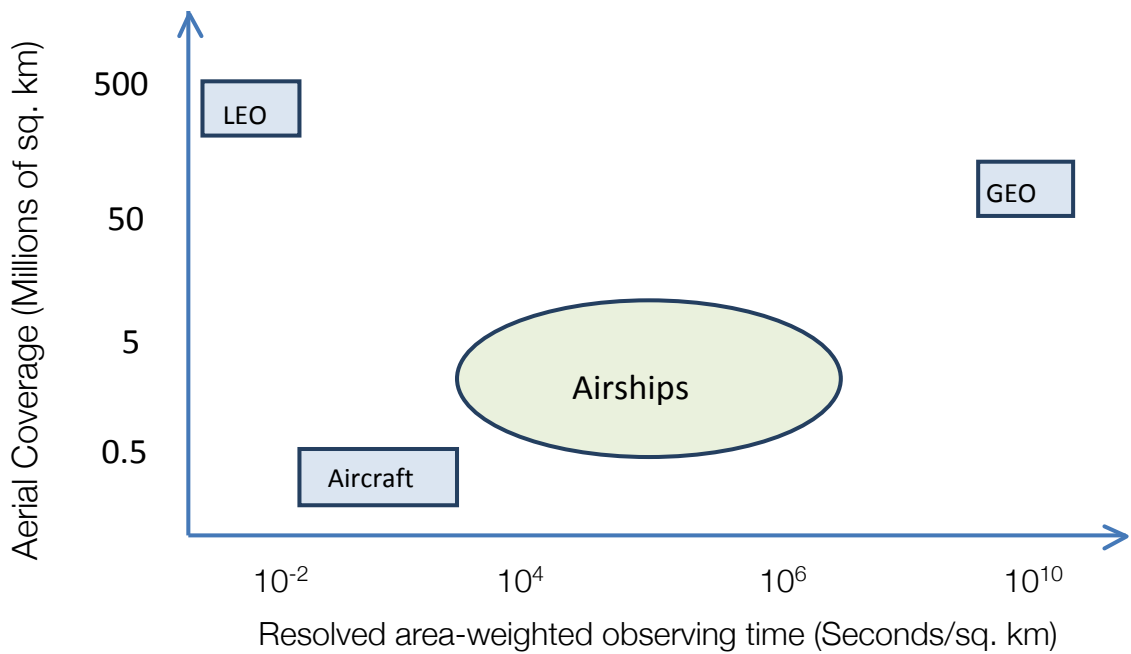

***Figure 1.1***: *Four regimes of Earth science measurement attributes:* **airship** *(high spatial and temporal resolution, diurnal to seasonal temporal coverage, local to regional spatial coverage),* **conventional fixed-wing aircraft** *(high spatial resolution, low temporal resolution, seasonal to inter annual temporal coverage, regional to continental spatial coverage),* **LEO satellites** *(moderate spatial resolution, low temporal resolution, weekly to inter annual temporal coverage, global spatial coverage),* **GEO satellites** *(low spatial resolution, high temporal resolution, diurnal to inter annual temporal coverage, continental to third-o-sphere spatial coverage).*

Attempts to study critical chemical and dynamical processes in the Earth's stratosphere are severely limited by available observing platforms.  Earlier investigations of stratospheric ozone utilized high altitude balloons and aircraft (ER-2, WB-57) for limited spatial (single location to hundreds of kilometers) and short duration (~4–10 hours) measurements.  As highlighted earlier in this section, there are currently many outstanding issues in Earth science that require longer duration observations at higher altitudes (e.g. stratosphere) beyond that possible with balloons and conventional aircraft.

For space sciences, fixed-wing aircraft, like SOFIA, can achieve multiple flights that add up to many hours of observing time, but come at great cost.  SOFIA is expensive to operate and can only do so for hours at a time, and can only reach maximum altitudes of about 45 kft under special circumstances (more typically operating at 39 kft). An absence of turbulence on an airship allows for greatly simplified telescope stabilization, instrument reusability, and increased duration of observations. Thus, airships as maneuverable lighter-than-air vehicles, are the most viable option for long duration, space-like stratospheric observations where the ability to station-keep is key.

## The Complementarity of Airships and Balloons

It is important to recognize that the use of airships for science is focused on those applications that existing free-flying balloon platforms either cannot meet at all or that provide substantially less attractive performance in terms of flight duration, reusability or geographic access. One set of such applications are those that require station-keeping over a particular geographic location. Free-flying balloons can at best provide only very short duration coverage in such situations during a brief overflight of the target area. Tethered balloons are a good solution to the station-keeping problem, but current technology limits the altitude of tethered balloons to approximately 15 kft. As a result, the need for long duration, high altitude station-keeping became a primary driver for the recent military investments in stratospheric airships.

Most free-flying balloons are limited to flight durations of up to a few days. This is particularly true of the stratospheric zero pressure balloons long used by NASA and other agencies for carrying large scientific payloads for astronomical research. An important exception to this limitation is Antarctic flight during polar summer where flight durations of 30-40 days have been achieved. However, the constant daylight in this environment precludes all optical astronomical observations and the geographic coverage is highly restricted.

NASA has been developing an Ultra Long Duration Balloon (ULDB) based on superpressure balloon technology with the objective of 3 month flights anywhere on Earth. However, this system is still under development and has yet to resolve the political obstacle that many countries refuse overflight permission for balloons. Given the prevailing stratospheric wind patterns, all long duration free-flying balloons will circumnavigate the planet leading to unavoidable overflight situations at any given latitude. The net result is that this promising technology may be severely limited in its operational endurance and geographic access, thereby leaving large gaps in coverage, especially in the northern hemisphere where the majority of the overflight issues exist. High-altitude balloon programs currently operate out of the following sites: McMurdo, Antarctica; Kiruna, Sweden; Alice Springs, Australia.  The lack of geographical deployment of balloons is predicated by its free-flying nature. An airship, under propulsion enjoys the scientific benefits of a stratospheric site along with the freedom of light-path control.

No free-flying balloon is reusable, although payloads are generally recovered after being parachuted to the ground. However, it is not uncommon for payloads to be damaged upon landing, particularly in remote geographic areas with hazardous terrain. Therefore, a significant advantage of an airship platform is the potential ability to bring the payload back to a safe landing at a specified location and thereby save the expense of repair or replacement of scientific payloads.

While high-altitude balloons have been and will continue to be a key asset for a variety of science goals, airships provide complementary abilities for missions requiring:

- **Increased payload capacity/flexibility** (ultimately, depending on the altitude, the promise is 500-10,000 lbs of payload and 100s of feet in length and/or width dimensions, needed for interferometric baselines)
- **Maneuverability** to follow or map phenomena
- **Ease and increased flexibility of payload launch and recovery**
- **Mission longevity** (weeks to months - possibly even years - rather than days)
- **Communications and data retrieval**

# Motivation for the Keck Institute Study on Science Aboard Airships

The original goals of this study were to identify and investigate the science capabilities of existing low-altitude airships, as well as stratospheric airships under development, to serve Earth, atmospheric, and space science goals. Airships as science platforms could have very different flight and duration characteristics compared to those for particular military or telecommunication missions. Scientists can be more flexible, adapting the ships' proven capabilities to their needs. However, while less stringent and narrow, the basic requirements of long-duration, mobility, and stability are similar.

Airships open up exciting new research opportunities. They could serve as complementary science platforms to both ground-based, other air-based and space-based facilities. Once reliable airships are fully developed, their advantages in greater duration and recoverability over existing high-altitude balloon flights in the stratosphere will make them a very powerful tool for scientific research across an unusually broad spectrum.

On September 5, 2012 the DoD released a statement in which it outlined its future aims for oversight and guidance for its continued investment in airship development (which includes various contractors participating in this study). These goals include a wish "to ensure cross-fertilization of [airship] technology", "coordinating interagency efforts", and "sponsoring airship-related conferences, reviews, table top exercises and academic studies as appropriate." This presents a timely opportunity to apply the detailed outcomes from this study to leverage significant multi-agency support for future follow-on studies and projects regarding a (multi-)science airship platform. Because of the great potential to create programmatic space, e.g. at the SMEX or MIDEX funding level for NASA (< $10 million USD), and equivalently at other relevant agencies, a key output of this study are the recommendations for both near- and long-term program strategies to be considered either individually by appropriate agencies or potentially towards combined, inter-agency efforts.

Momentum for a comprehensive study of science opportunities on airships has grown over the last several years. Some study members have even helped to lead this activity: Fesen has been researching airships for over 10 years and has organized workshops on the use of stratospheric airships for astronomical observations. Miller led communication with the airship team at Northrop Grumman to explore the parameters of developing various airship science platform from 2010 until the commencement of this study. In April of 2012, Kauffmann and Goldsmith organized a successful workshop to discuss both balloon and airship opportunities for space and earth science at JPL. Presentations were given by members of both JPL and industry (including Lockheed Martin and Northrop Grumman) with several participants of this study in attendance, including Lord who has developed a sophisticated atmospheric model to project observing conditions from various airship altitudes. Rhodes, who has had a continued interest in high-altitude observing, organized a NASA task-force workshop in July 2012 regarding communication and data retrieval for high-altitude balloons and airships. Duren has led various efforts in the last several years to exploit existing lower-altitude airships for Earth and atmospheric science. Due to the increasing interest apparent from these events, actions and the discussions which followed, the leaders of this study organized a systematic program to clearly define what science can best utilize these platforms and how, and importantly, to chart a practical way forward with key experts and leaders gathered from their respective fields.

Over the course of the study new goals were generated. (1) We assessed ways to spur the development of stratospheric airship technology to maturity so that it may become a viable platform for use by the science community, and (2) we assessed the ability of tethered balloons to meet the needs of science, including concepts of new stratospheric tethered platforms.

# Chapter 2:
# Past, Current, and Future Airships

## Defining an Airship

Airships are lighter-than-air vehicles that can generate their own thrust for maneuvers. The external structure or envelope of such craft fall into three categories: rigid, semi-rigid, and non-rigid.

Rigid airships are built with material stretched on a full structural frame. The Zeppelins are an example, using light alloy girders to support gas cells. Semi-rigid craft typically have a lower arced keel that maintains the ship's shape. Non-rigid ships have envelopes that are supported by the gas pressure alone (e.g., the blimps). Airships currently under study include the so-called "hybrid" designs that not only rely on buoyancy from helium to stay aloft, but also incorporate an aerodynamically shaped airfoil that provides additional lift. Some new designs additionally employ "vectored thrust", where the on-board propulsion may be aligned in any direction, offering additional altitude control. In the hybrid airship, the three forces of buoyancy, lift, and thrust are combined to direct the craft.

## A Brief History of Airships

The earliest airships were French inventions comprised simply of manned balloons carrying hand-powered propulsion devices. The first, in 1784, used a hand-cranked propeller, while a second, a year later, using wing-like structures, successfully crossed the English Channel. Airship builders soon employed the latest technological advances with designs progressing rapidly during the second half of the 19th Century. Propulsion systems were soon powered by steam engines, internal combustion engines, and even electrical motors. During this period, war spurred-on the development of airships, chiefly for military surveillance. In a wide variety of designs, the virtue of these craft in providing surveillance and reconnaissance capabilities (so long as the platform required only modest mobility) was repeatedly demonstrated. The first decade of the 1900's marked the early use of airships for scientific investigations as Wellman and Vaniman attempted, albeit unsuccessfully, exploration of the North Pole from airships.

The new century saw worldwide enthusiasm for the use of airships with many important developments: the mass production of the successful non-rigid body Astra-Torres airships by the French; the advent of the successful Zeppelin series of rigid-body airships; the emergence of numerous production companies in France, Italy, and Germany, and the Goodrich Company in the US, and the first trans-Atlantic crossing attempt.

WWI was a turning point for the use of airships. While the Germans expended enormous effort and resources in a fleet of airships capable of bombing and machine-gunning, it was soon realized that their actual effectiveness was relatively

*Airships: A New Horizon for Science*                                                                 13

insignificant. The ships were too prone to mechanical failure, bad weather, and general bombardment inaccuracy. These airships terrorized but did not inflict significant damage from a military perspective. While the hydrogen-filled ships were (contrary to belief) not explosive and leaked only slowly when punctured by bullets, they were quite susceptible to devastating fire from incendiary projectiles – fires, which the hydrogen lifting gas did in fact aid. Incendiary counter-attacks effectively ended the bombing role of airships. But what was learned during the war was the airships' value for surveillance. They made good scouts, which was especially useful for naval war vessels approaching distant shores. After the war, treaties forbade further German military airship development.

The period between WWI and WWII marked the heyday of the large rigid airships with examples including Germany's commercially oriented Graf Zeppelin and Hindenburg and the US Navy's Akron and Macon. Many of these airships were lost in crashes, most famously the crash of the Hindenburg at Lakehurst, New Jersey on May 6, 1937. The rest were dismantled early in WWII. Development also continued on non-rigid airships between the wars to the point where they were used in WWII for antisubmarine patrol missions. After the war, larger versions were used for anti-submarine patrol and airborne early warning missions, thanks to the addition of a large radar mounted inside the hull. The last of these non-rigids were decommissioned in 1962.

The US Navy attempted to get back into airships in the mid-1980s with the YEZ-2A Naval Airship Program. This was to be a large, radar-carrying airship for airborne early warning missions, Westinghouse–Airship Industries (WAI) won the contract with their Sentinel 5000 design. As part of the program, a half scale demonstrator, the Sentinel 1000, was constructed and flown. While funding for the Naval Airship Program was cut in 1989, the Sentinel 1000 continued development and flight testing until it was destroyed in a hangar fire in 1995.

## Airships Currently in Use or Under Development

Among the results from the tragedy of September 11, 2001 and the resulting Global War on Terror has been a renewed interest in airships and their potential in affordably providing a platform for persistent intelligence, surveillance, and reconnaissance (ISR) missions. The relatively permissive air defense environment in Iraq and Afghanistan coupled with the need to follow insurgent movements over lengthy times to counter such things as improvised explosive devices fit well with airship capabilities. Several examples of the different kinds of DoD programs are shown in Figure 2.1 and Table 2.1.

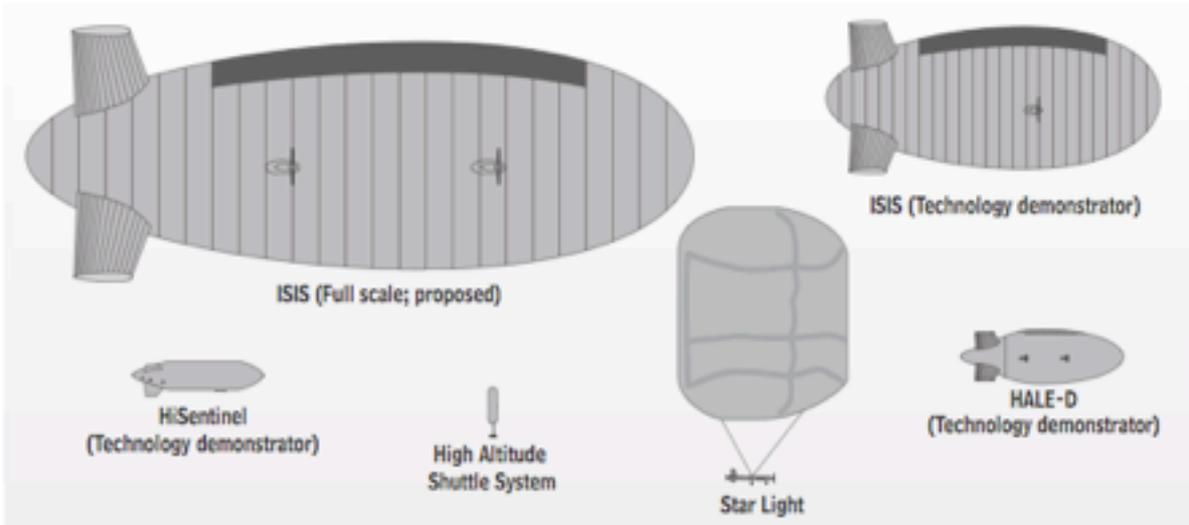
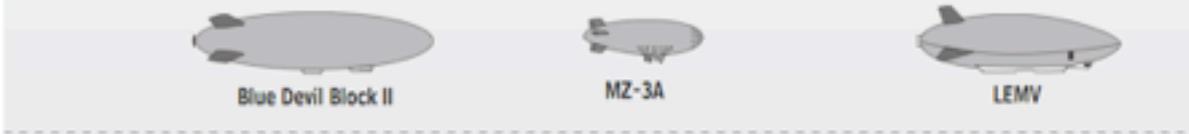
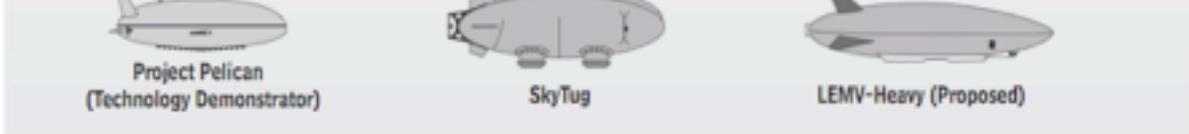
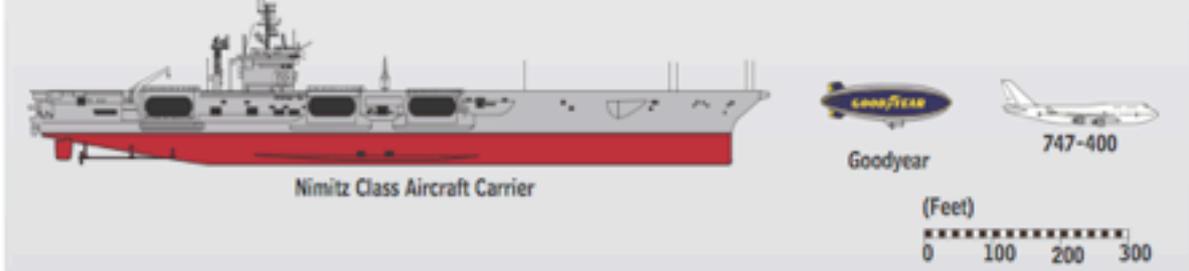

*Figure 2.1 (Exhibit 1 of the of the Congressional Budget Office's 2011 report: "Recent Development Efforts for Military Airships.") This illustration shows that current airships under development range in size from that of the Goodyear Blimp to an aircraft carrier. (The "airlift" or transportation airships shown above, are not germane to this report.)*

| Table 2.1: General Characteristics of Airships Under Consideration by DoD (2011) | | | | | | |
|---|---|---|---|---|---|---|
| Mission | Operating Altitude | Airship Type and Examples (1st Flight) | Altitude (Feet) | Endurance or Range | Status of Technology | Characteristics |
| Intelligence, Surveillance, & Reconnaissance | Low Altitude | Conventional  BD2 (2012) MZ-3A (2006) | Up to about 20,000 | 100 to 300 hours | One system currently operating; Others under construction | Relatively mature technology |
| | | Hybrid  LEMV (2012) | Up to about 20,000 | 500 hours | Technology demonstrations ongoing | Uses static lift from helium, aerodynamic lift from the shape of the envelope, and vectored thrust to stay aloft |
| | High Altitude | Conventional  HALE-D (2011) HiSentinel (2005) ISIS (2010) | 65,000 to 75,000 | Greater than 400 hours | Technology demonstrations ongoing | Very large envelope volume to sustain lift |
| | | Payload-Return  Star Light (N/A) | 65,000 to 75,000 | 100 to 300 hours | Technology demonstrations ongoing | Payload is detachable and returns to the point of origin; airship is single-use |

**Table 2.1** Based on Exhibit 3 of the Congressional Budget Office's 2011 report: "Recent Development Efforts for Military Airships." This table shows the various airship classifications recently and currently under examination under various DoD programs. Several examples are also shown. More information is given in the following section.

## Current Operational or Planned Airship Examples

### BD2

The Blue Devil 2 airship, built by Mav6, is a conventional non-rigid designed to fly at 20 kft for 4 to 5 days with a 2,500 lb ISR payload including onboard processing that makes it an aerial data fusion node. Originally scheduled for first flight in the fall of 2011, the program was cancelled in June, 2012 due to technical and programmatic challenges.

### MZ-3A

The Navy's MZ-3A, a modified American Blimp Corporation A-170 commercial airship, is a 178 ft long non-rigid ISR airship that carries a crew. It is currently the only operational airship in the DoD and is used for payload test and evaluation. It was used recently to monitor the Deepwater Horizon oil spill in the Gulf of Mexico. The MZ-3A is a platform for up to 2,500 lb of cameras, radar and other sensors. It flies at up to 9.5 kft and cruises at 40 mph. Its typical flight duration is 10 hours but it has a 24-hour capability.

### LEMV

The LEMV is a non-rigid airship of hybrid design. It was developed for deployment in Afghanistan in 2012. It can operate at 20 kft for up to 21 days an can produce up to 16 kW of electrical power and carry a 2,500 lb ISR payload. Schedule delays and weight growth reduced the altitude to 16 kft and flight duration to 16 days by the time the first flight was performed in August, 2012. The program was cancelled in February of 2013 and the vehicle was deflated and sold back to its builder, Hybrid Air Vehicles (HAV) in October, 2013.

### HiSentinel

The HiSentinel program is a family of high altitude airships to provide persistent communications and ISR capability to the DoD. The HiSentinel program was a tactical airship demonstration program for the DoD to demonstrate the various key technologies for a stratospheric airship. The HiSentinel systems were comprised of the airship, ground support systems, weather support system, and flight/payload command, control and communications ground station. Six high altitude airship engineering flights have been conducted over the years with five of those flights achieving greater than 65 kft altitudes. All key stratospheric airship technologies were demonstrated during the development program.

### HALE-D

The HALE-D is a high altitude conventional non-rigid airship demonstrator for the HAA, the larger High Altitude Airship. Intended to operate at 60 kft for two to three weeks with a small demonstration payload, the first flight occurred in July, 2011. Unfortunately a problem occurred during ascent and the flight was terminated after rising to only 32 kft. The airship came down in a heavily wooded area of southeastern Pennsylvania. During recovery operations, the hull caught fire and was destroyed. Funding for the program ended in 2011.

### ISIS

The ISIS (Integrated Sensor Is Structure) Demonstration System Program is a conventional non-rigid airship that includes an integrated Radar system. The airship is 511 feet in length and operates at an altitude of 65 kft for one year. Originally intended for a first flight beginning in late 2012, cost and technical challenges have caused the program to delay airframe development and to refocus on radar risk reduction testing, which will complete in mid-2014.

### Star Light [3]

The U.S. Navy's Naval Air Warfare Center awarded a Phase 1 and Phase 2 contract to begin development of a next generation stratospheric airship with a radically new design. The vehicle, named StarLight, was proposed to deliver unprecedented performance in operating altitude, flight duration and forward velocity. The uniquely designed vehicle would supposedly operate at 85 kft above the earth's surface powered solely by photovolatics. Current status appears to be inactive.

## Science vs. Military Requirements for Airships

The modern development efforts toward advanced airships and drones in the US have been largely driven by defense initiatives and requirements. Comparisons between drone technology and airship technology have included such attributes as the visibility of the craft, its radar cross-section, and its thermal and acoustical signatures. These concerns are absent when designing airships for science and therefore the designs may be better optimized for the tasks at hand and the costs may be further reduced.

The advantages that airships hold for science in some cases are the very disadvantages that apply to military applications. As an example, immobility. Being relatively slow-moving platform can be a liability for defense, but a very desirable attribute for instance in astronomy – critical in gaining highly accurate gyro-control to maintain telescope pointing. The relatively relaxed propulsion constraints for science reduces one of the biggest stumbling blocks in the development process for the military. Part of the military concern about the viability of airships includes their inability to be defended. So as to enhance the efficacy of high altitude airships, modern design efforts have gone into making the airships harder to detect by reducing their acoustic and thermal signatures and using an envelope material that will absorb radar frequency radiation. Since science applications obviate these concerns, further design optimization can occur for airships used for science applications alone.

The trade-off between the use of satellites and airships to provide space-like observing environments is of concern to both the science community and the military. Perhaps the most important consideration is cost. Historically, satellite observatories cost between 0.1 and 3 billion USD, while airships of the full-scale ISIS are more in the order of 10-100 millions of USD. (We note that stratospheric airship cost estimates are currently unreliable, since their development is so new.) The smaller-scale airships can cost on order of a million dollars. Unlike satellites, airships do not typically require persistent and costly global-scale control, tracking, and communication. They can operate more typically on a local scale. Finally, for most airships, the payload of the experiment may be changed-out and the ship reused.

---

[3] http://www.globalnearspace.com/press_release_SPSA.shtml

# Chapter 3: The Stratospheric Environment

## Summary

The advantages of placing an airship at stratospheric heights, namely space-like observing conditions, unavoidably lead to space-like challenges for the airship and the instrumentation on board. The environment more closely resembles outer space than sea level: low temperatures, low air pressure, higher solar irradiance, and lower convective heat transfer all pose challenges to the reliable operation of instruments. These challenges are not intractable, however: we know how to build instruments that operate in space and in the near-space conditions of balloon-borne platforms.

Rather, the most significant challenge posed by the stratospheric environment is to the airship platform itself. Harvesting the solar energy needed to station-keep against the stratospheric winds can be difficult at some times of year and in some geographic locations. However, the stratospheric wind environment can be quite benign at some latitudes and during the summer months, and the "sprint and drift" approach to airship navigation can significantly alleviate the challenges of airship propulsion.

Propulsion is the dominant source of power consumption in a stratospheric airship, where the power needed to station-keep scales as the cube of the wind velocity. As we show in this chapter, the winter winds at high latitudes are significantly less benign than the summer winds, though this seasonal difference is much more pronounced in the southern hemisphere. This indicates that the power requirements for a stratospheric airship can be minimized by operating either in the summer hemisphere, where conditions are extremely benign (perhaps moving back and forth between the hemispheres as the seasons change), or in the tropics where winds are slightly stronger than in the summer hemisphere and in some cases more variable but conditions are relatively constant throughout the year.

In this chapter we describe the stratospheric wind environment, thermal environment, and solar illumination that determines the power available to an airship. Finally, we will discuss some of the challenges in powering the airship propulsion and instruments.

## Challenges of the Stratospheric Environment

The environment at the desired altitudes of operation for stratospheric airships (60-70 kft) more closely resembles outer space than sea level conditions. The environment can pose operational problems for most off-the-shelf systems. All flight hardware must be carefully selected and tested to ensure that it will function properly in a stratospheric environment (Table 3.1) (Smith, 1999), particularly when integrated with other heat generating equipment. Some of the issues that need to be considered when developing equipment for stratospheric flight are:

- The air is very cold, averaging -55° Celsius.

- The air pressure is low, 3-5% of sea level pressure.

- The air density is low, 15-20 times lower in density than sea level air density.

- Convective heat transfer is reduced significantly, so radiative heat transfer mechanisms are needed.

- Solar irradiation can be 25 - 37% higher than at sea level.

- UV radiation is more intense than at sea level.

- Exposed surface temperatures can drop as low as 194 Kelvin (-79° Celsius) at night, due to radiation cooling.

- Getting rid of waste heat generated by electronic systems is a significant concern, and is driven more by radiation than convective cooling.

- Electrical arcing is more likely in a low density atmosphere, and single event upsets can be more common if the electronics are not designed properly.

| Table 3.1: Global Flight Environment | |
|---|---|
| **Atmospheric:** | |
| Tropics: | -90C @ ~ 50-60 kft altitude |
| Polar: | -45C @ ~ 30-35 kft altitude |
| Mid-Latitude: | -55°C @ ~ 45-60 kft |
| | -80°C in summer |
| | (seasonal & latitudinal fluctuations) |
| **Radiation:** | |
| Solar Constant (seasonal) : | 1358 W/m$^2$ (nominal) |
| | 1312 W/m$^2$ (minimum) |
| | 1404 W/m$^2$ (maximum) |
| Albedo : | 0.1 (minimum) |
| | 0.9 (maximum) polar |
| Earth Flux: | 90.7 W/m$^2$ (minimum, Tropospheric cloud top temperatures of 200°K) |
| | 594. W/m$^2$ (maximum, Desert @ 320°K planet temp.) |

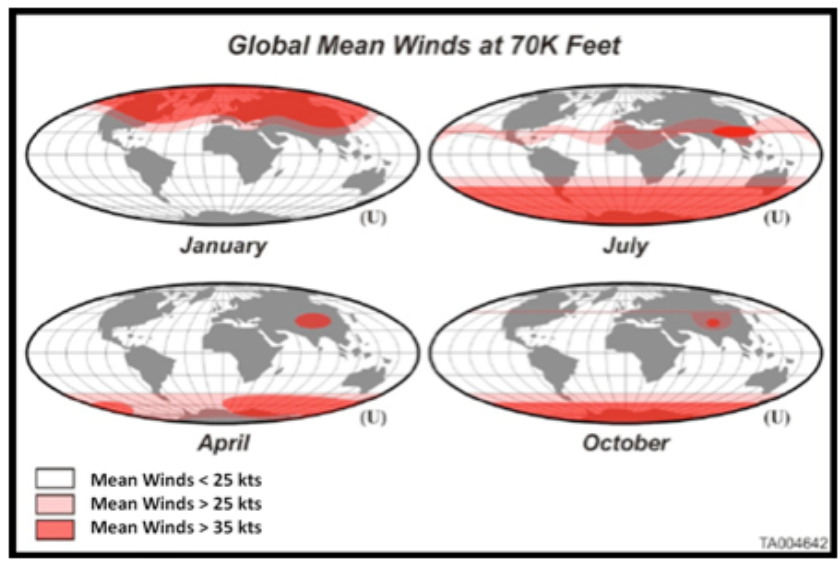

*Figure 3.1*: *The stratospheric wind environment is dependent on geographic location and time of year (Perry 2002).*

## Stratospheric Winds at 20 km (~65 kft)

The stratospheric wind environment is highly dependent on geographic location and time of year, as shown in Figure 3.1 and Figure 3.2. It is dominated by the zonal wind component (East-West winds, u, Figure 3.3), which is characterized by a strong cyclonic (westerly) circumpolar jet in the winter hemisphere (mean winds of 20-30 m/s at 65 kft) and weak (~5 m/s) easterly winds in the summer hemisphere. The transition from easterly to westerly (or vice-versa) in the extratropics occurs over a period of a few weeks during the equinoctial seasons (March-April-May and September-October-November).

The lower stratosphere (~65 kft) is optimal for stratospheric airship flight because the winds in this altitude range are statistically the slowest. It should be noted that the wind speed magnitude and the altitude at which the minimum wind speed occurs vary with geographic location and time of year. Figure 3.1 (Perry, 2002) illustrates the geographic and seasonal variation of wind speed at a particular altitude (Jaska, 2005). Generally, the altitudes for minimum winds are from 60–70 kft (50 to 40 millibars).

In the tropics, the zonal wind at 20 km alternates between easterly and westerly with a period of ~27 months; this is known as the stratospheric quasi-biennial oscillation (QBO). The easterly phase of the QBO is stronger than the westerly phase, with mean 20 km equatorial easterly winds of 15-20 m/s compared to ~5-10 m/s during the westerly phase. The meridional component of the wind (North-South winds, v, Figure 3.4) is much weaker than the zonal component, with 65 kft mean values <1 m/s at all latitudes throughout the year. Vertical velocities ($\omega$=Dp/dt, where p is pressure) are on average less than 0.25 cm/sec at 65 kft.

Wave motions at a range of spatiotemporal scales from < 10 meters in < 1 hour to > 1000 kilometers in > 1 month play a critical role in the stratosphere, and in fact drive the mean meridional overturning circulation (Brewer-Dobson circulation, e.g. Haynes et al. 1991). These waves, which include gravity waves (whose restoring force is gravity), synoptic- and planetary-scale Rossby waves (whose restoring force is planetary vorticity), tides (gravity waves modified by rotation and compressibility), and mixed Rossby-gravity waves, result in large deformations of the stratospheric flow and thus large deviations of u,v,w, and other parameters such as temperature from their mean values.

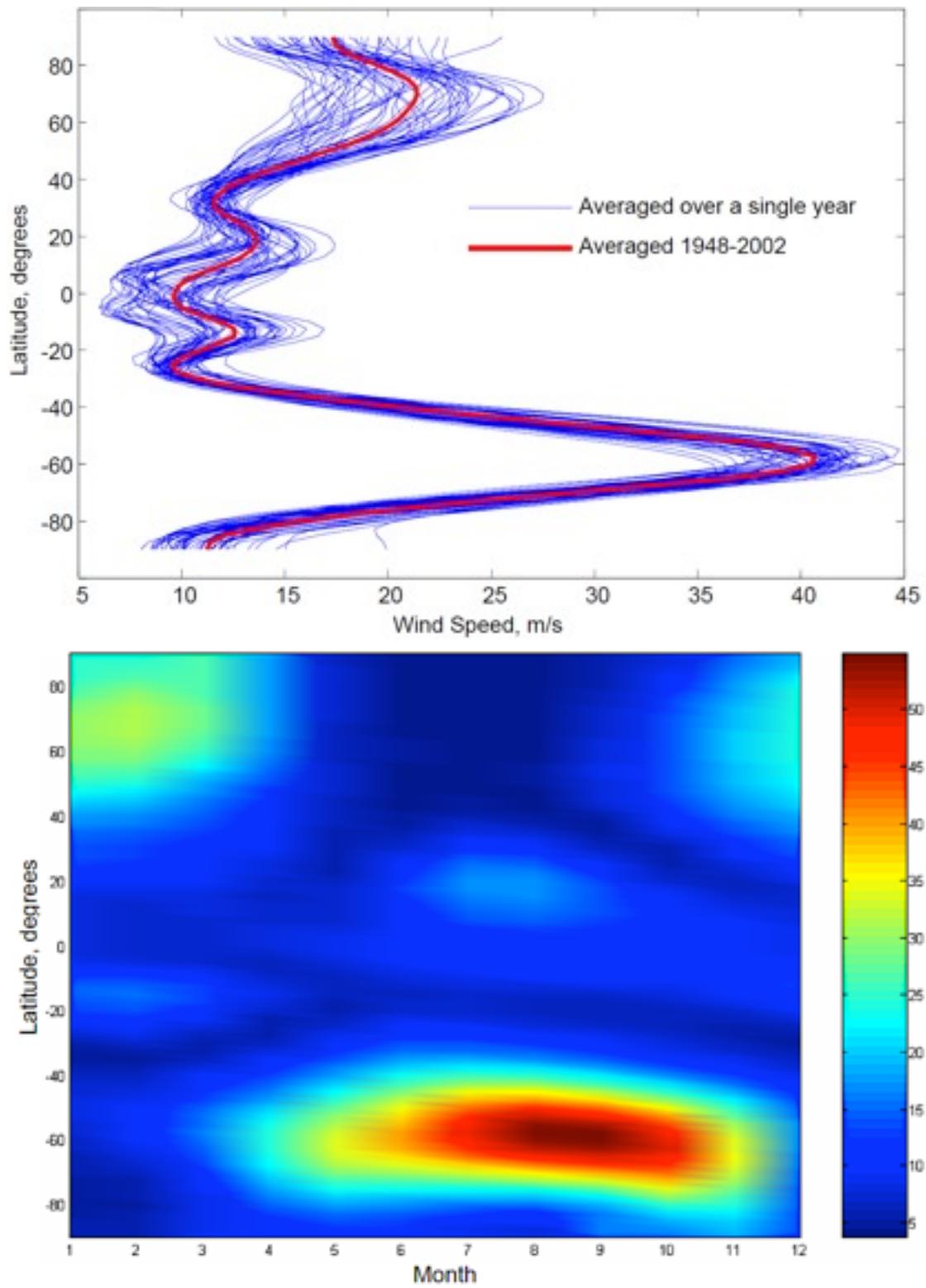

***Figure 3.2***: *Stratospheric winds in the antarctic are more severe than elsewhere on the planet, particularly in the southern winter. (Jaska 2004).*

## Zonal winds (u)

The zonal wind in the extratropics is characterized by a seasonal cycle, as well as a hemispheric asymmetry in the winter circumpolar winds. The southern hemisphere westerlies are much stronger than those in the northern hemisphere because there are fewer land masses in the southern hemisphere to generate waves, which disturb and exert a drag on the mean zonal flow.

These differences are illustrated in Figure 3.3, which shows the variability of the zonal wind amplitude u from 3-hourly output of the Modern Era Retrospective Analysis for Research and Applications (MERRA) reanalysis system (Rienecker et al. 2011) for 2005, divided into 6 latitude bands. The reanalysis output is generated by the Goddard Earth Observing System Data Assimilation System Version 5 (GEOS-5, Rienecker et al. 2008), which utilizes a wide variety of atmospheric observations to constrain a high resolution atmospheric model.

In addition to the seasonal cycle in the mean wind, Figure 3.3 also shows a pronounced seasonal cycle in the variability of the wind in the extratropics. Variability is much higher in the winter than in the summer, because Rossby waves, which make up a large portion of the stratospheric wave spectrum, propagate only on westerly winds (Charney and Drazin, 1961). During the summer, when the mean wind is easterly, Rossby waves evanesce above the tropopause and thus have little impact on the flow at 20 km. Furthermore, the source strength for gravity waves, which can propagate on both easterly and westerly winds, is weaker during summer. In the tropics, the zonal winds are weaker than in the extratropics and do not show a pronounced seasonal cycle in either mean values or in variability. The QBO transitioned from westerly phase to easterly phase at 20 km during 2005; this is reflected in the mean wind values, particularly in the 0-30N latitude band.

## Meridional winds (v)

While the mean meridional wind is < 1 m/s at all latitudes throughout the year, its variability is much larger than that of u in the extratropics during winter; the 5th and 95th percentile values of v can exceed 40 m/s, rivaling the zonal wind speeds. This is illustrated in Figure 3.4. If the airship is oriented east-west, presenting a large surface area to the meridional wind, these large fluctuations of v during winter are of particular concern. During summer (as well as in the tropics throughout the year), the variability in v is much weaker, and wind speeds rarely reach ±10 m/s. Variability of the vertical velocity, ω, is similar to that of v, with values reaching as high as ± 20 cm/s (compared to a mean value of < 0.25 cm/s) during winter and much smaller variability in the summer and in the tropics (max|ω|<3 cm/s).

**Long-term variability:** In addition to analyzing the variability of the 3-hourly MERRA output, we also examined the interannual variability of 20 years (1991-2010) of monthly mean wind speeds. In all cases, the variability of the three-hourly output exceeds the interannual variability (IAV), though for u the two are comparable. The winter hemisphere short-term variability in v is more than double the IAV, and for ω the short-term variability exceeds the IAV by a factor of 20. The large difference between variability determined from 3-hourly output and that determined from monthly mean output demonstrates the critical role that waves play in generating stratospheric variability.

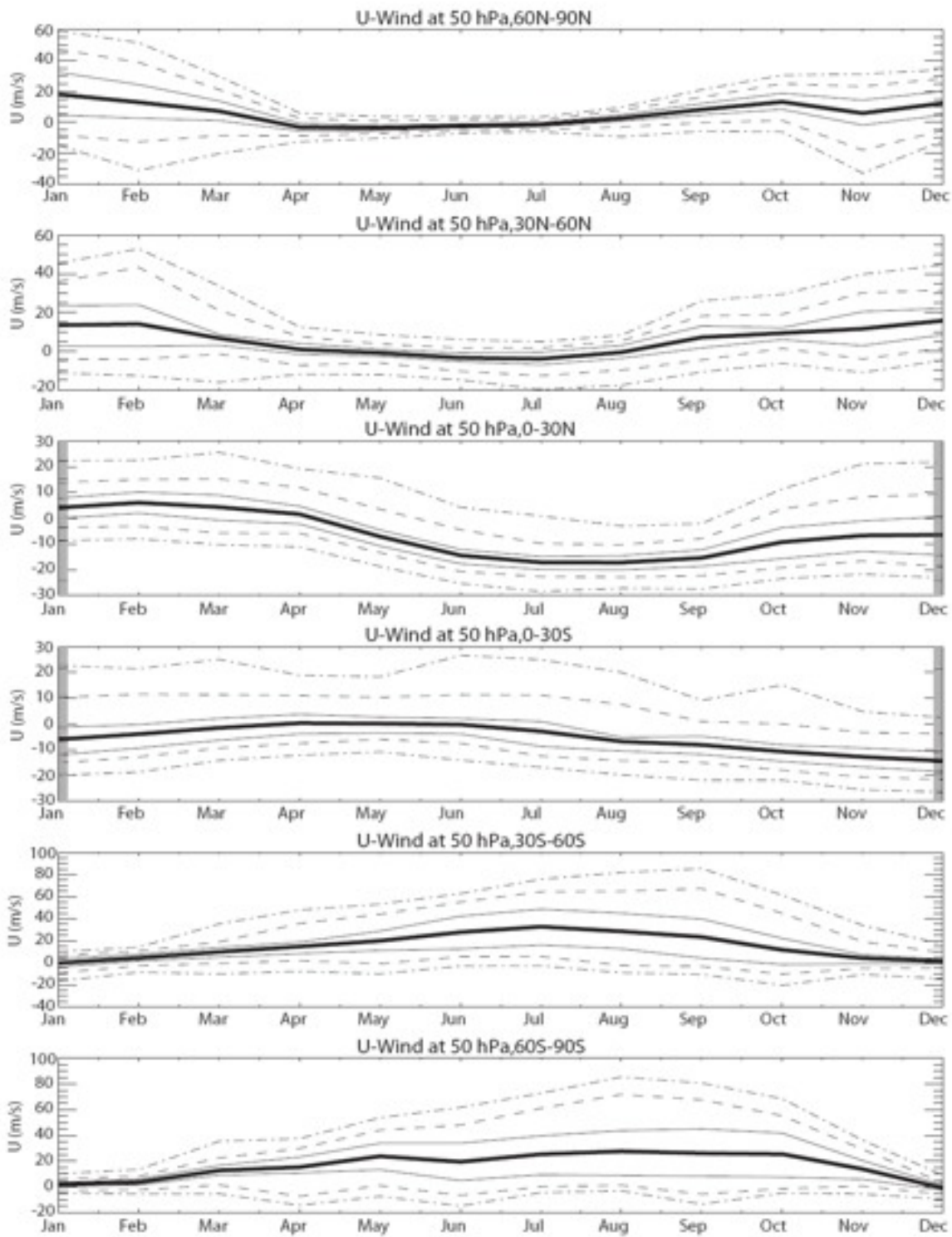

***Figure 3.3***: *The temporal and spatial variation in the zonal component (u) of the stratospheric winds, from MERRA. The lines show mean (thick), 25th/75th percentiles, 5th/95th percentiles, min/max.*

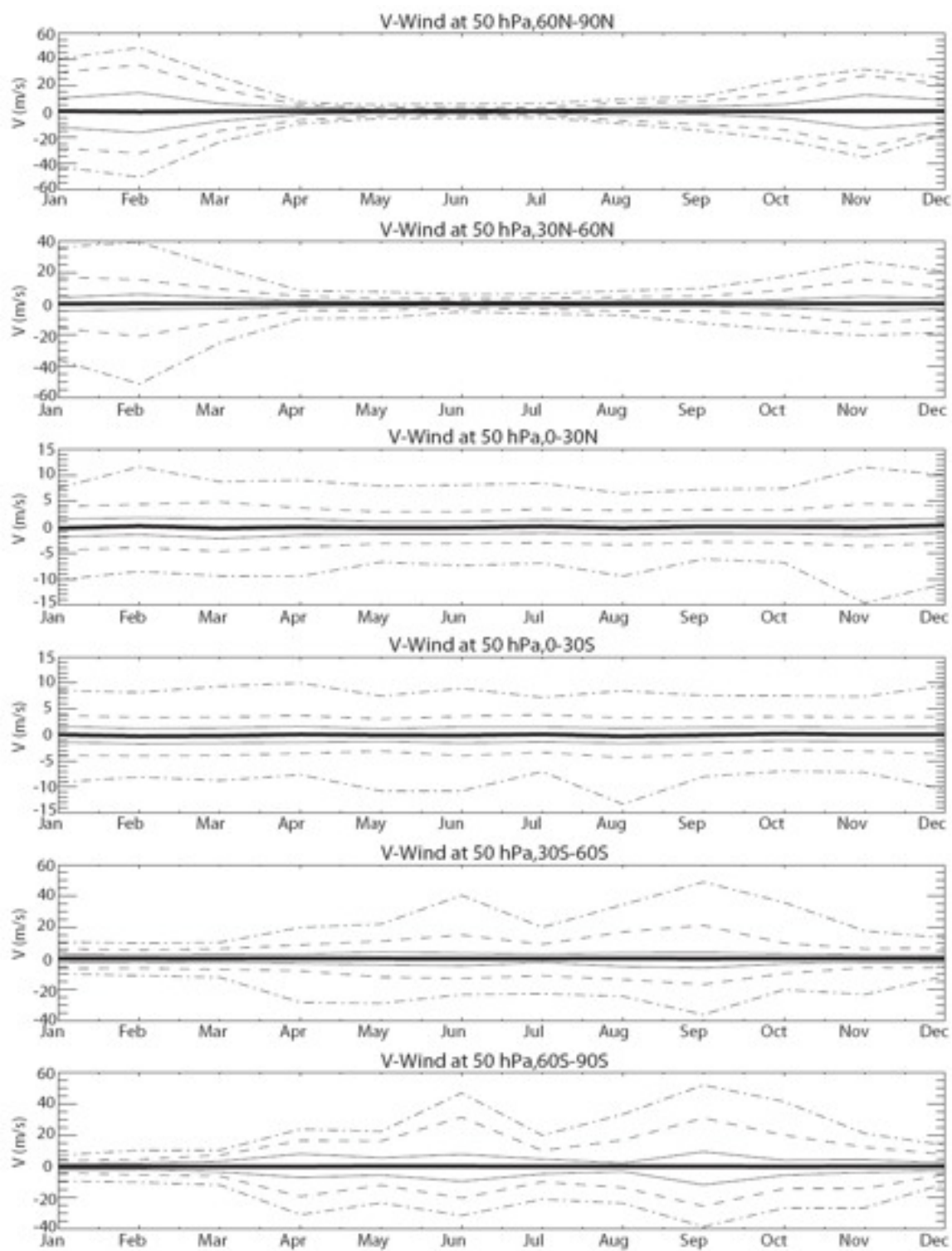

***Figure 3.5***: *The temporal and spatial variation in the meridional component (v) of the stratospheric winds, from MERRA. It is much weaker than the zonal component (u, Figure 3.3). The lines show mean (thick), 25$^{th}$/75$^{th}$ percentiles, 5$^{th}$/95$^{th}$ percentiles, min/max.*

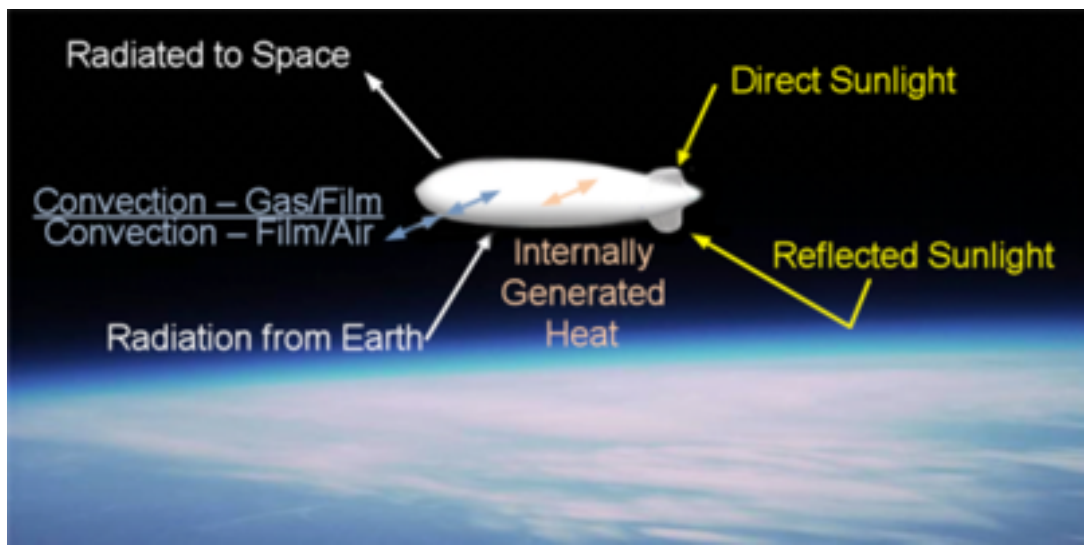

*Figure 3.5*: An illustration of the heat loads on an airship in the stratosphere. Credit: SwRI/HiSentinel

## Thermal Environment at 65 kft

The stratospheric airship is a thermal vehicle, just as is any free-floating balloon, as illustrated in Figure 3.5. As such, the key to a successful stratospheric airship is managing the thermal balance of the vehicle's lifting gas. Low altitude airship technology generally cannot be used in long endurance stratospheric airships since it does not address the thermal balance problem resulting from the near-space environment.

Thermal and structural models are used in the design process of the concepts. The thermal model is used to predict the temperature extremes that a stratospheric airship might experience. The mechanical model is used to design the stratospheric airship hull to ensure it is strong enough to contain the differential helium pressures, to maintain hull pressurization, to ensure there is enough propulsion power to achieve the desired air speeds, to calculate the masses and distribution of all airship components, and to ensure there is sufficient lift to carry the mass of everything on the stratospheric airship to cruise altitude. The design process is an iterative process with information being exchanged between the thermal model and the mechanical model. The stratospheric airship radiative properties and preliminary convection coefficients are inputs to the thermal model.

The atmospheric temperature and its variability can be assessed using MERRA, as illustrated in Figure 3.6. The atmospheric waves (Rossby waves, gravity waves, etc) that affect the stratospheric winds also affect the stratospheric temperatures. The difference between the monthly maximum and minimum temperature in the winter extratropics can exceed 50 K. In summer, the variability is very small at the poles and is typically ~±10 K in midlatitudes (as well as in the tropics throughout the year).

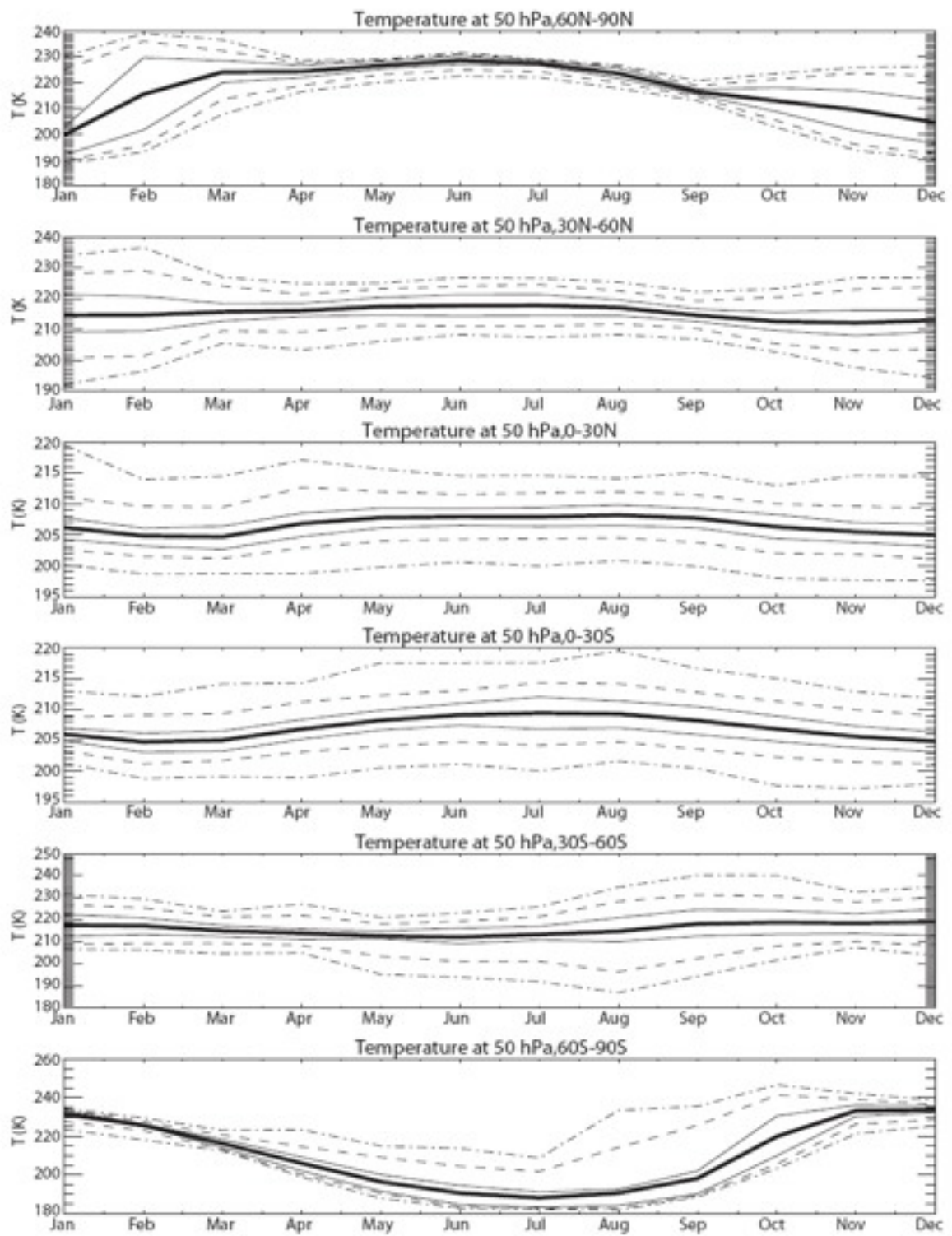

***Figure 3.6***: *The stratospheric atmospheric temperature spatial and temporal variation, from MERRA. The lines show mean (thick), 25th/75th percentiles, 5th/95th percentiles, min/max.*

# Solar Environment at 65 kft

Power to a stratospheric airship is supplied by the mounting of photovoltaic (PV) arrays within or on the surface of the airship. As such, the solar availability is of crucial importance to an airship. Figure 3.7 shows that the solar availability for a mid-latitude site (Washington DC) varies from ~9.5 hours in mid-winter to more than 14 hours in mid-summer, with a more extreme variation closer to the poles.

A portion of the energy gathered by the solar arrays is stored in rechargeable batteries or regenerative fuel cells to fuel night-time operation. Support circuitry is available to completely manage the charging and discharging. The battery control circuitry must be designed to eliminate the possibility of the energy storage system going off line unexpectedly as well as controlling charging rates, levels and power shedding.

Mounting of PV arrays on the hull surface introduces some inefficiencies due to the curvature of the hull thereby increasing overall system mass, requiring a larger airship. It also introduces large temperature variations due to the daytime heating of the arrays that the hull material must be thermally isolated from so as to not cause hull structural failure.

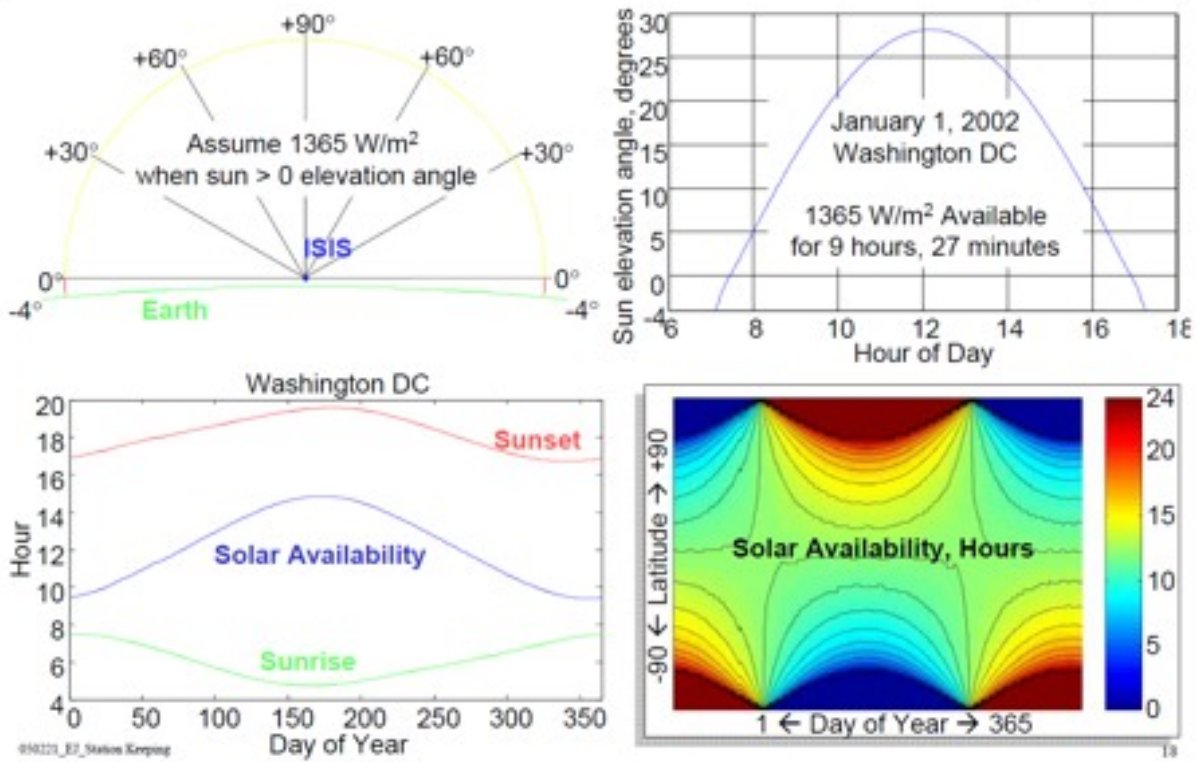

*Figure 3.7*: The solar availability for a mid-latitude site, such as Washington DC, varies from ~9.5 hours in mid-winter to over 14 hours in mid-summer. (Jaska 2004)

# Airship Propulsion

Stratospheric airship power requirements include the power needed to supply the sensor payload, all support systems and in particular the propulsion system. The propulsion system power will be by far the largest driver for power and is a function of the wind speeds at the cruise altitude.

Vehicle propulsion power requirements are a function of velocity and are proportional to the air speed raised to the third power. Therefore the power required to either cruise to a target location or the power required to station-keep over a target location is given by the expression (Eq. 3.1):

$$\text{Power}(vel) := \frac{\rho_{air}}{2 \cdot \varepsilon_{prop}} \cdot \text{Vol}_{hull}^{\frac{2}{3}} \cdot C_d \cdot vel^3 \qquad \text{(Eq. 3.1)}$$

where:

$\rho_{air}$ = air density

$\varepsilon_{prop}$ = propulsion efficiency (motor, transmission, propeller)

$\text{Vol}_{hull}$ = hull volume

$C_d$ = drag coefficient based on shape and Reynolds number

$vel$ = velocity

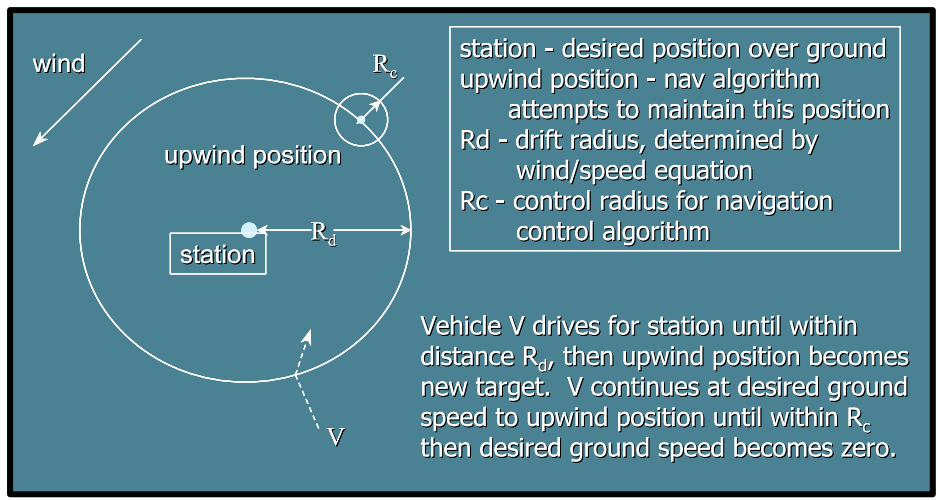

***Figure 3.8***: *Sprint and drift station-keeping navigation. Credit: SwRI/HiSentinel*

**Sprint and drift:** A major factor affecting airship size is the mass associated with the power generation and energy storage systems required for airship propulsion. The efficiencies of the solar cells and the energy storage system are very important. Propulsive energy mass minimization is one consideration when selecting the operational navigation modes of "station-keeping" or the "Sprint and Drift" approach. Figure 3.8 shows how the stratospheric airship flight controller may implement the "Sprint and Drift" approach for station keeping.

**What does "Sprint and Drift" mean?** Assuming that the average wind and average airship speeds are equal, the airship sprints upwind of the station keeping point during the day at high speed, and during the night, drifts back over and then downwind of the station keeping point at a slower nighttime speed. This technique can significantly reduce the total mass for the propulsion power system. For some airship designs, the "Sprint then Drift" technique reduced the propulsion power mass by 33% below an airship of equivalent volume that could achieve the same speed at night as during the day.

**How does the "Sprint and Drift" approach save weight?** For an example airship design, it takes 9.1 grams of equipment to produce 1 watt of power during the day. It takes 48.2 grams to produce 1 watt of power from the fuel cell system for a 14-hour night. It is advantageous from a mass minimization standpoint to spend a little more energy during the day in order to conserve power during the night. The calculation is not simple, but for these airship designs, the minimum mass is achieved with a night-to-day speed ratio of 0.46, thus to achieve an average air speed of 15 m/s for a 10 hour day and a 14 hour night, the day speed is 21.9 m/s and the night speed is 10.1 m/s.

## Atmospheric Turbulence

Atmospheric turbulence can be described by the Fried parameter $r_0$, which describes the size of telescope that can achieve diffraction-limited seeing. At ground level it is typically 10 cm (an average site) to 20 cm (an excellent site). It is a measure of the seeing of a site, which is the degradation of image quality due to the perturbation of wavefronts by a turbulent medium.

Scintillation Detection and Ranging (SCIDAR) measurements, e.g. Hoegemann et al 2004 in Proc SPIE v. 5490, show that $r_0$ can be >1.5 m at 65 kft. This means that telescopes as large as 1.5 m can achieve diffraction-limited seeing in the stratosphere. This assumes that the airship-equivalent of dome seeing, i.e. the turbulence in the atmosphere produced by the airship itself, can be mitigated.

# Chapter 4:
# Case Studies for Science Aboard Airships

*This chapter highlights particular case studies which exploit some of the more unique features of airship platforms. These cases are merely a subset of the wide range of science made possible on airships.*

## In Earth and Atmospheric Science

The near- and long-term applications of airships for science span a wide range of science questions. The ability of airships to station-keep above a given spot on the Earth for continuous regional monitoring of Earth science processes (atmospheric variations, landmass changes, coastal biology, etc.) may be the most compelling single application.

As detailed in recent International Climate Assessment reports (e.g. IPCC, 2007), changes to our Earth system are expected to continue, driven in the future by periodic natural events (e.g., volcanoes, solar cycles) and human emissions resulting from evolving technology, population distribution, land use, and international agreements. These changes will influence global climate and temperature, as well as the precipitation that affects the availability of fresh water that we drink, the build-up and transport of pollutants that affect the quality of the air we breathe, and the health of ecosystems that provide the food that we eat.

Study participants identified three Earth system processes where breakthroughs in understanding can be achieved using airships, namely:

- Megacity emissions and air quality,
- Tropical carbon cycling, and
- Coastal ecosystems.

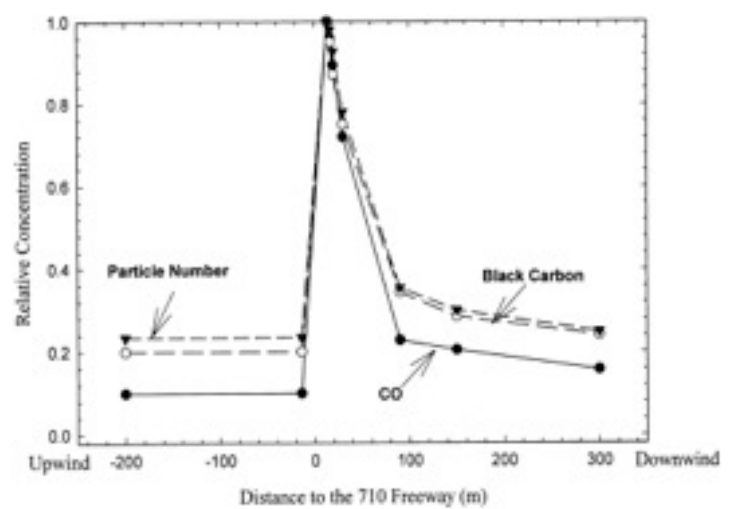

*Figure 4.1*: *Localized particulate and carbon monoxide concentrations associated with a major freeway, from Zhu et al, 2012.*

## Megacity Emissions and Air Quality

### Background

Twenty three cities in the world contain more than 10 million people each and are termed "megacities" (Figure 4.3). These megacities, while relatively small in spatial extent, are responsible for a significant fraction of the world's emissions of ozone and aerosol precursors and greenhouse gases. Within these environments, the emitted ozone and particulates have serious health impacts and are responsible for many deaths each year. Health impacts strongly depend on cumulative exposure to pollution, where exposure times vary on scales of a few hundred meters and depend on the temporal characteristics of the emission sources (e.g. strengths and directions, see Figure 4.1).

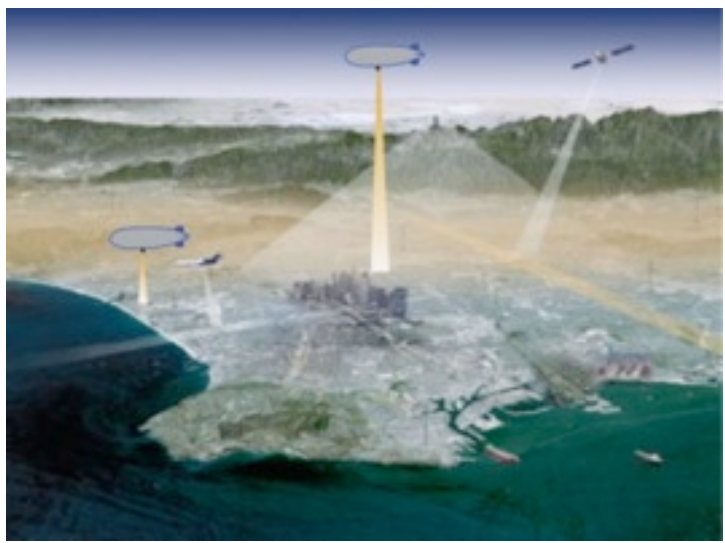

*Figure 4.2*: *Megacities concept from Duren and Miller (2012) with high- and low-altitude airships added to complement satellite, aircraft, and ground-based observations.*

As highlighted by Duren and Miller (2012), a focus on megacity emissions allows the characterization of a sizeable fraction of anthropogenic global carbon and ozone-precursors within a vastly smaller observational footprint. The scientific challenge is to quantify pollutant and greenhouse gas emissions and understand how the emissions, transport, and chemistry interact to produce variations in the quality of air contained in the "urban domes" surrounding the megacities (Figure 4.2).

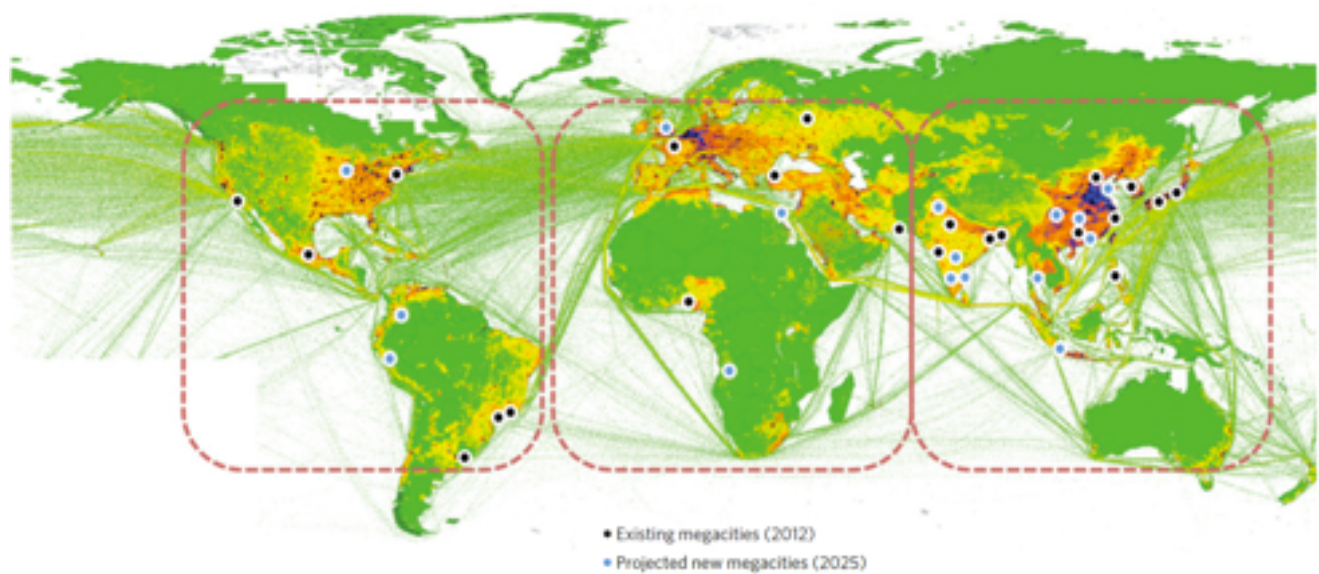

*Figure 4.3*: *Megacity locations, from Duren and Miller (2012)*

**Requirements**

In order to benefit greenhouse gas monitoring and health impact assessments, the characterization of megacity urban domes requires observations of ozone and aerosol precursors and greenhouse gases at high spatial resolution and with near surface sensitivity for long durational periods over the full extent of the megacity. Remote sensing ultraviolet/visible/near-infrared hyperspectral and multispectral spectrometers have sufficient spectral resolution to meet the chemical species observing requirements. The typical specifications for these instruments are 200 kg (mass), 150 W (power), and 0.2 $m^3$ (volume). With respect to the combined observational platform and instrument, the requirements are:

1) >$10^5$ $km^2$ FOV

2) <10 meter spatial resolution

3) > 2 day duration

4) >Monthly repeat

**Approach**

Required measurements would be accomplished by positioning an airship above select megacities for extended periods. In order to obtain the needed FOV and duration, the preferred platform would likely be a stratospheric one. However, trade-offs between airship altitude and speed, duration and repeat could make other airship (and tethered aerostat) options viable.

The airship approach provides substantial cost and performance advantage over the GEO satellite solution offered by Duren and Miller (2012). In particular, the GEO orbit can only enable horizontal resolutions of order 4 km, given reasonable telescope sizes and detector performance at GEO.

## Tropical Carbon Cycling

**Introduction**

As the largest reservoirs of above-ground carbon in the world (Figure 4.4), tropical forests play a substantial role in the global carbon cycle. Recent studies estimate that nearly 70% of the terrestrial sink resides in tropical forests (Pan et al. 2011). However, uncertainties in these estimates are large owing to the very small number and scale of observational samples and the statistical bias inherent in the sampled data. Uncertainties in flux measurements are even larger and leave unclear whether tropical forests are net carbon sources or sinks. Fortunately, new methods are emerging for quantifying carbon uptake and respiration in forests such as lidar and radar based biomass change detection and spectrometer-based measurements of solar-induced fluorescence (as a proxy for gross primary production, GPP) and $CO_2$ fluxes (Figure 4.5). With new instrument tools nearly in-hand, the major impediment to be addressed is achieving large enough sample areas with the requisite spatial resolution and near-surface sensitivity.

*Airships: A New Horizon for Science* 33

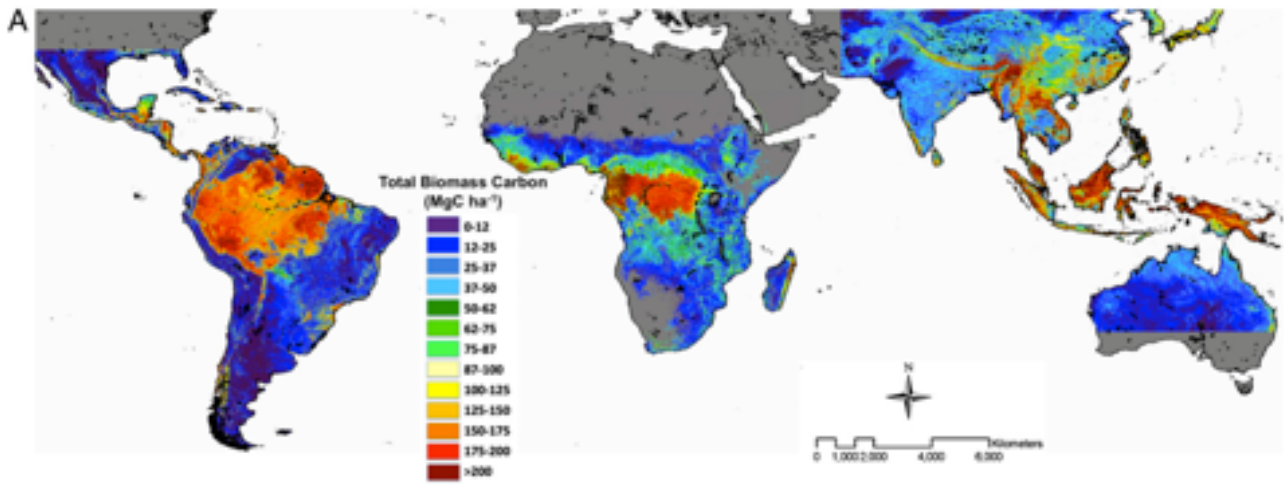

*Figure 4.4*: Benchmark map of tropical forest carbon stock, from Saatchi et al. (2011)

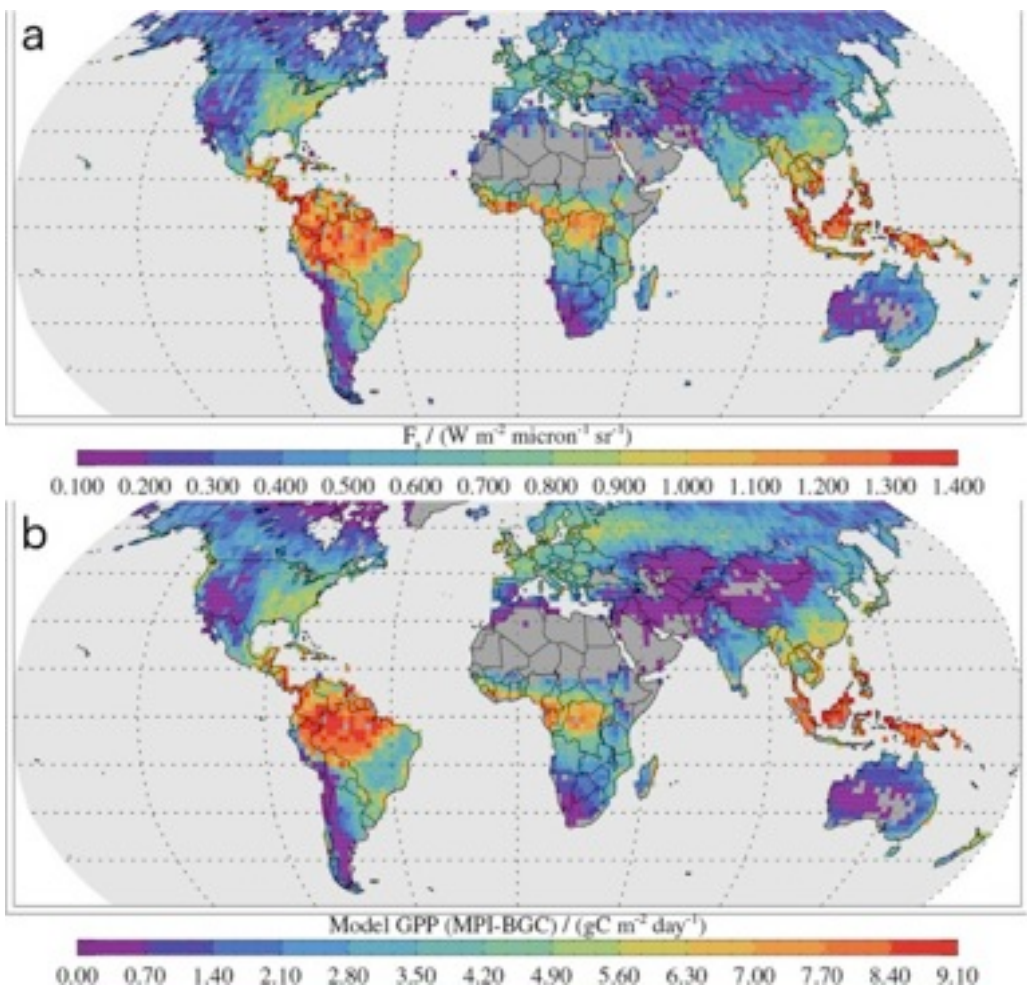

*Figure 4.5*: Chlorophyll fluorescence maps, from Frankenberg et al (2011)

**Requirements**

Characterization of the net flux of carbon within tropical forests requires relatively long-term, albeit periodic measurements of forest growth, degradation and regrowth as well as atmospheric $CO_2$ gradients across the forests.   Visible and near-infrared lidars and L-band and P-band radars are required for retrieving tree height information. Multispectral spectrometers are required for inferring net carbon exchange rates from GPP and $CO_2$ concentrations.  On the one hand, a full set of instruments would constitute a substantial payload, namely, 500 kg of mass, 1kW of power, and 0.2 $m^3$ of volume (plus a 0.5 m diameter antenna).  On the other hand, a mission focused only on global primary production and CO2 concentrations would require substantially less platform resources of order 200 kg mass, 150 watts power, and 0.2 $m^3$ volume.  With respect to the combined observational platform and instrument, the requirements are:

1) >$10^6$ $km^2$ FOV

2) <10 meter spatial resolution

3) > week duration

4) >Seasonal repeat

**Approach**

Required measurements would be accomplished by traversing an airship across the Amazon (or other tropical forests) for multiple days to collect a large, statistically representative biomass sample.   As with the megacity case, the needed FOV and duration is best met with a stratospheric platform.  However, the large resource requirements of the full instrument suite represents a severe challenge for the airship development and suggests that other lower altitude airships and tethered aerostat options be considered.

## Coastal Ecosystems

**Introduction**

The areas of land and sea surrounding coastlines are home to a vast array of plants, animals and nutrients that provide a myriad of goods and services to human societies (Figure 4.6).  They are also home to an increasing fraction of the world's population, with over 40% of people living within 150 km of a coastline.  The adverse impacts of human activities and climate change on the health of coastal ecosystems are gaining increasing recognition and scrutiny (Halpern, 2008).  In particular, scientific attention is focusing on alteration of runoff and coastline topography as well as alteration/destruction of coastal habitats. Particularly vulnerable habitats include salt marshes, mangroves, and coral reefs. In order to unravel the complex interactions between the many ecosystem components, a persistent, multifaceted observational approach is needed for characterizing the coastal zone processes. A strong, first step in this direction would be to examine the relationships between ecosystem health, large-scale coastal ecosystem drivers, such as river runoff flow and tidal activity, and changes in basic habitat structure (e.g. river and ocean topography and bathymetry).

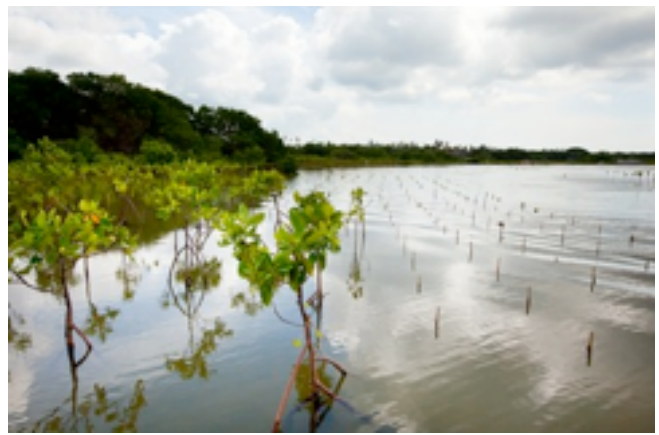

*Figure 4.6*: Coastal wetlands (Marc Simard)

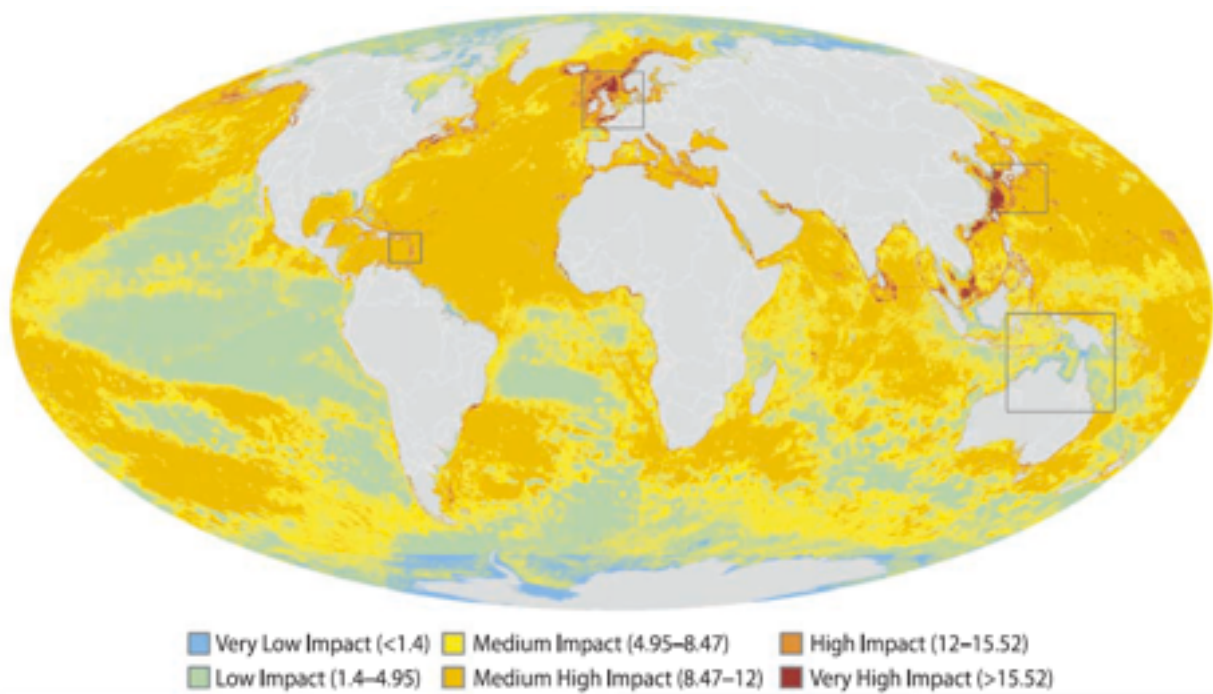

*Figure 4.7*: Cumulative human impacts on various ocean ecosystems, from Halpern et al. (2008)

**Requirements**

Diagnosing relationships between ecosystem drivers, physical habitat changes and general ecosystem health requires frequent coastline observations over a full tidal cycle. The required observations would include ocean currents, coastal water levels (i.e. river and sea heights), river and ocean bathymetry, and ecosystem function/ocean color (as a proxy for health). A full instrument suite consisting of a K-band radar (for current), visible lidar (for topography and bathymetry), and a multispectral imaging spectrometer for ecosystem function would require payload capabilities similar to the tropical carbon case described above, namely, 500 kg of mass, 2kW of power, and 0.2 m$^3$ of volume (plus a 0.5 m diameter antenna). A threshold requirement would be for measurements of ecosystem function/ocean color at select areas with substantial in situ instrumentation (e.g. height gauges). Platform resource requirements would be substantially reduced for the threshold mission to 100 kg mass, 1500 watts power, and 0.2 m$^3$ volume. With respect to the combined observational platform and instrument, the requirements are:

1) >10$^5$ km$^2$ FOV

2) <10 meter spatial resolution

3) > 2 day duration

4) >monthly repeat

**Approach**

The optimal approach for this case would be to position a nearly stationary airship above the selected coastal area for days to weeks at a time. As with the tropical carbon cycling case, the resource requirements for any of the coastal measurement ideas represent a severe challenge for high altitude airship development. Consequently, options for alternative lower altitude airships and tethered aerostats should be considered.

# Astrophysics and Planetary Sciences

Space sciences have a long history of groundbreaking ballooning projects, and as described in Chapter 1, airships' unique capabilities may be best suited as complementary to traditional high altitude balloons. Many astronomers, solar physicists and planetary scientists desire to be above nearly all of the atmosphere, either to reduce atmospheric absorption of signals, or to avoid turbulence that degrades imaging capabilities. While airships envisioned today do not achieve altitudes of 120 kft to reach certain regimes of UV and X-ray wavelengths, in theory a tethered balloon could potentially achieve this altitude. Rather stratospheric airships offer a complementary benefit of extremely long observation times at mid-latitudes (i.e., with night-time observing in the UV, visible, and near infrared), without the risk of data loss if a crash were to occur over an ocean. Additionally, the significant size of a full-scale airship allows for a compelling new approach to interferometry with long baselines above most of the atmospheric effects.

Atmospheric absorption of radiation particularly hampers astronomical observations in the infrared through sub-millimeter wavelength regimes. The absorbing species responsible are among the most abundant molecular species in the Earth's atmosphere, including: $H_2O$, $O_3$, $CO_2$, $N_2O$, $CO$, $CH_4$, and $O_2$. Above 40-60 kft, where the tropopause begins, most water vapor will condense out into liquid or ice form, and atmospheric transmission dramatically increases. While water vapor is not the only absorber, it is a dominant one, and observations conducted above this altitude begin to see a clearer transmission window.

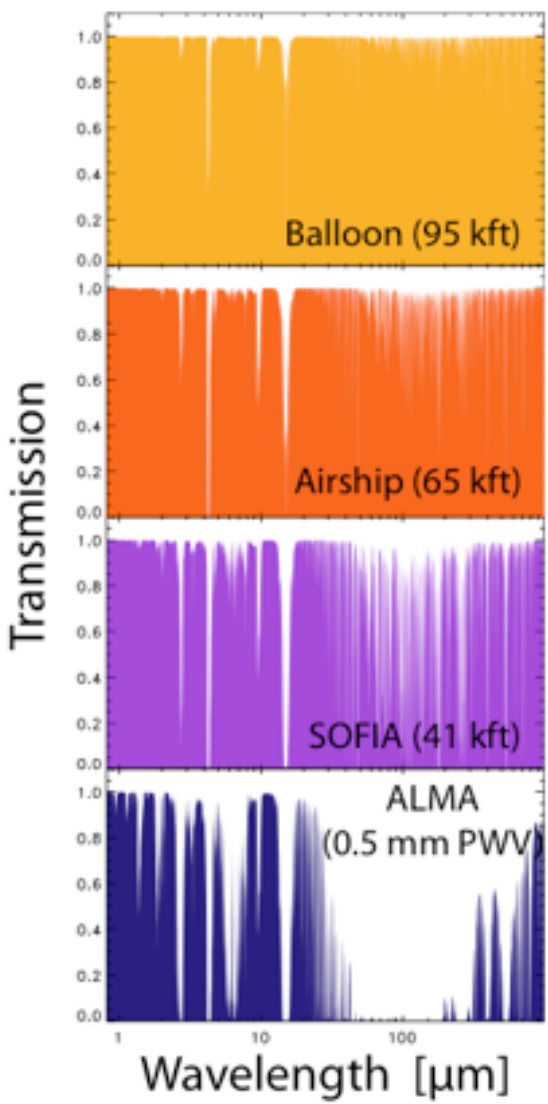

*Figure 4.8: (left) shows the modeled atmospheric transmission between 0.8 micrometers and 1 millimeter based on the ATRAN (Atmospheric TRANsmission) software (Lord 1992). Shown is the transmission at 4 altitudes: the ground-based ALMA Observatory at 16.6 kft, SOFIA at 41 kft, an airship at 65 kft and a balloon (or airship) at 95 kft. The overhead ozone in each case is $3.8 \times 10^{19}$ molecules per $cm^2$ and the perceptible micrometers of water vapor is 0.5, 7.3, 1.1, and 0.16 respectively. Note that ground-based observatories are practically unable to observe from 30 to 200 micrometers, and at 41 kft, from SOFIA transmission coverage is broken apart by numerous narrow gaps. Above 65 kft the transmission becomes near unity across these bands, allowing for astronomical access to the far-infrared and sub-millimeter spectral windows.*

## Airship-based Sub-mm Interferometer for Imaging Black Hole Event Horizons

**Introduction**

Airships may be an ideal platform for Very Long Baseline Interferometry (VLBI) observations at sub-mm wavelengths. The major limiting factor for ground-based sub-mm VLBI observations is phase noise introduced by the atmosphere: it leads to decoherence of the signal that fundamentally limits the sensitivity of the observations.

A solution to the problem of atmospheric decoherence is to get above the atmosphere. A fleet of orbital sub-mm satellites would achieve this, but faces significant funding and technological challenges. A fleet of stratospheric airships may be cheaper and more flexible than a satellite mission.

The principle scientific aim of this project is direct imaging of the event horizon around black holes. The event horizon of the black hole at the center of M87, a nearby giant elliptical galaxy, has an angular size of about 20 μas. The resolution of an interferometer in arcseconds is given by $2.1 \times 10^5 \lambda/D$. Here D is the baseline distance. Very roughly, VLBI observations with baseline lengths of 5000 km at a frequency of 500 GHz will resolve the event horizon. The baseline length between Hawaii and the US East Coast, for reference, is approximately 8500 km. There is an existing project attempting the same measurement using ground-based facilities called the Event Horizon Telescope.[4] It currently uses the SMA in Hawaii, CARMA, and a telescope in Arizona, but will soon include ALMA. An airship pathfinder or fleet to follow up on the ground-based facilities could provide better placement of antennas and less de-cohering atmosphere to look through at the higher frequencies (say 500 GHz) that will give better resolution. There are some significant engineering challenges to doing VLBI at 500 GHz though, most importantly having good reference clocks.

**Requirements**

A proper reckoning of the viability of this idea will require more careful analysis, but here are some basic requirements based on interferometry rules-of-thumb:

**Stability:** The minimal specification on the stability of the platform is that the distance between any two antennas needs to be stable to a fraction of a wavelength (~1/20th) during the acquisition of one data sample. If we sample the voltage waveform at 1 GHz, this requires that the distance vary by less than 30μm in 1 ns (assuming a 500 GHz observing frequency). This translates to keeping the baseline stable to within 30m in 1 ms. The station keeping requirements will be set by whatever technical solution is chosen for measuring the absolute position of each antenna.

**Storage/telemetry:** If the data are acquired at a rate of 1 GHz, and assuming 4 bits per sample, the data rate which needs to be stored/transmitted at each antenna is ~480 MB/s. This is the minimum data rate. It is possible that multiple frequency channels will be digitized and transmitted.

**Positional accuracy:** If a source is bright enough, we don't need to know the absolute position of each antenna. As long as the fringes can be detected in less time than the decoherence timescale, the effects of changing baselines can be modeled and corrected for. For dimmer sources we need to know the absolute position of each antenna. The positional accuracy will be similar to that of the stability: we need to know the position of each antenna to better than 1/20th of a wavelength.

---

*4 http://www.eventhorizontelescope.org/*

**Large Scale Option**

- 10 airships at baselines between 500 and 5000 km (can be confined to US airspace if necessary).

- Each airship has a single ~4m class antenna on it (can be lightweight carbon fiber/composite antenna).

- Each antenna is equipped with a 500 GHz MMIC-based heterodyne receiver which digitizes the incoming waveform.

- Digitize a 30 GHz bandwidth and integrate samples down to the stated rate of 1 GHz.

- Wirelessly transmit the resulting data stream to storage hardware on the ground, or if the data rate is too high write directly to disk on the airship.

- The position of each airship is measured by triangulation from base stations on the ground using something like laser range finders combined with GPS.

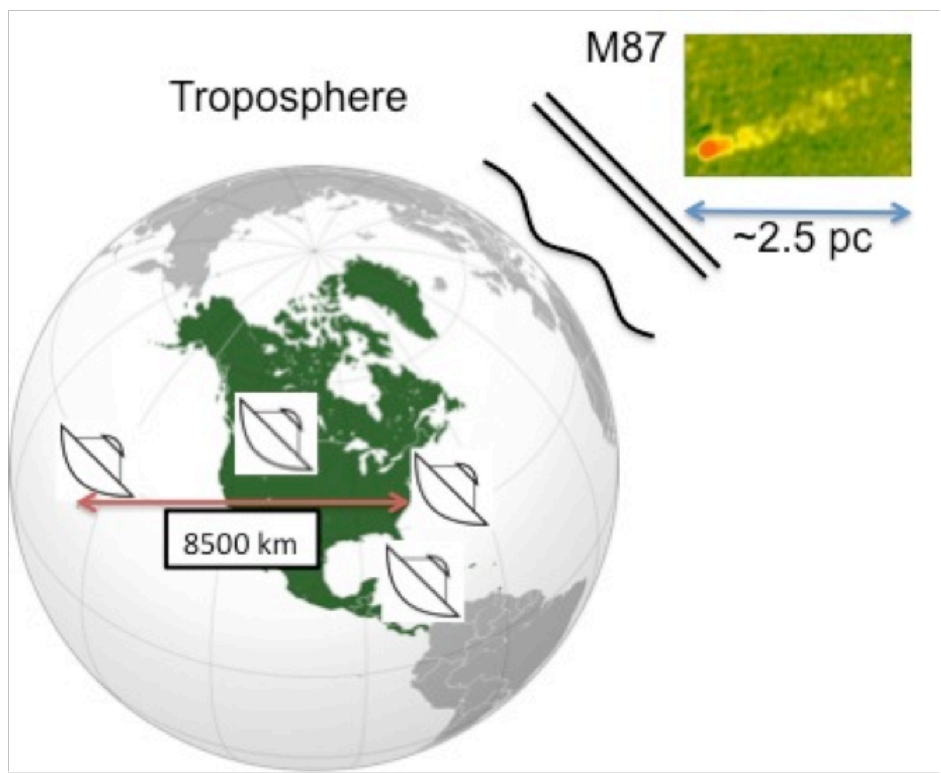

*Figure 4.9*: *A schematic representation of VLBI imaging of black hole event horizons. A network of airships spread across the US can have baselines up to 8500 km, providing an extremely high resolution (at 500 GHz) of 1 milliparsec of the environment around the supermassive black hole M87 - 60 times higher than the best imaging resolution to date shown here (M87 image credit: NRAO). Stratospheric airships would be above most of the distorting water vapor in the atmosphere, which is concentrated in the troposphere and below.*

**Pathfinder Mission**

Before a mission like the one proposed in the previous section can even be contemplated we need to demonstrate that interferometry from an airship is technically feasible. Here we outline a pathfinder mission to demonstrate this.

- Launch a single airship equipped with a ~3m class antenna and a ~300 GHz receiver.

- The waveform will be digitized and wirelessly transmitted to a ground station to minimize the payload mass.

- Simultaneous observations will be made of bright radio sources with the airship receiver and a receiver on a ground-based telescope such as the CARMA or the SMA.

- We will correlate the data from the airship receiver and the ground-based antenna and attempt to detect fringes.

This mission will demonstrate whether an airship is stable enough for sub-mm interferometry and will allow us to test technical solutions for monitoring the absolute position of the airship antenna.

## Sub-mm/THz Interferometer for Protoplanetary Disk Science

Another area of sub-millimeter, or TeraHertz (THz), astrophysics in which airships may provide an ideal platform are interferometric studies of the protoplanetary disks encircling young stars. While many molecular species can be studied from ground-based aperture synthesis arrays such as ALMA, light hydrides, especially water, remain out of reach due to the low atmospheric transmission from even dry mountaintop observatories. The delivery of water to young terrestrial planets is a key step in the development of habitable worlds, and so measurements of the so called "snow line" in protoplanetary disks - that distance at which water vapor begins to condense - and how it varies with time and the properties of the pre-main sequence star are central to exoplanetary science.

Studies above the water vapor in the atmosphere are the obvious solution, and single telescopes that can observe the universe at infrared through THz wavelengths such as Spitzer, Herschel and SOFIA have made important contributions to the detections of water vapor emission lines from protoplanetary disks. Typical distances for the snow lines around Sun-like stars are some 1-3 Astronomical Units (AU), or about 7-20 mas at the distances of the nearest star-forming clouds. The strongest and most diagnostic water vapor rotational emission lines from disks occur between 30 and 300 microns, which would require telescopes of nearly a kilometer in diameter to provide a direct image of water vapor on such scales. Fortunately, the ordered Keplerian velocity field in disks provides exacting radial information on the molecular emission observed, provided the line shapes are accurately measured and the orientation of the disk is known (from scattered light or mm-wave observations of dust and CO, for example). The challenge, then, is to provide a platform that can detect the faint water vapor emission from the innermost regions of the disk at high spectroscopic resolution.

The coherent nature of even a single baseline THz interferometer provides just such a platform, and when placed at airship altitudes could study water vapor and numerous additional species. Such an instrument cannot be placed on SOFIA, and airships may provide a cost-effective alternative to multiple THz telescopes in space.

**Requirements**

While a more detailed study will be required to determine the actual collecting area and detector sensitivity required, a few requirements follow from basic interferometry:

***Baselines/Stability:*** Here, the baseline need only be up to a few tens of meters to resolve out any extended emission from the molecular cloud. Ideally the baseline should be stable to fractions of a wavelength over timescales of minutes to hours, but if necessary metrology could be used to measure and correct for baseline fluctuations in the electronics that feed the backend.

***Receivers:*** Sensitivity will be critical, and so superconducting heterodyne receivers will likely be greatly favored over semiconductor-based devices (such as MMICs). High efficiency Stirling coolers would then be needed to provide the necessary cryogenic conditions at the focal plane for a long duration mission.

***Backend/Telemetry:*** A pair of high speed digitizers followed by an FPGA to provide real time averaging is all that would be required for the correlator, and with sufficient pointing and baseline stability the data could be coherently averaged for minutes. Thus, the data rates should be quite modest compared to the sub-mm Event Horizon telescope(s), even with hundreds of thousands of spectral channels. The power requirements are fairly modest for such a system, which can be housed on a single card.

**Pathfinder Mission**

There is great synergy here with the Event Horizon pathfinder concept. With a single ~3m class antenna and sub-aperture illumination of two detectors, the single baseline interferometer concept could be fully tested (except for the independent telescope mounts). Water masers can be chosen as bright beacons for such a mission, several are available in the 300-600 GHz region.

**Large Scale Option**

Here, the minimum requirements would be two 4-10m class antennas (they could be lightweight carbon fiber/composite structures) on a fixed baseline. High THz transmission fabrics should be tested for airship environments to determine whether the antennas can be placed inside the airship envelope, which would be optimal for stability. With three telescopes, the three independent baselines provide for phase closure, and would be a significant enhancement over a single baseline system (and could be distributed linearly along the long axis of the airship).

# Hubble-Competitive Imaging

**Diffraction-limited imaging at optical wavelengths**

At stratospheric heights the atmospheric turbulence is low enough that diffraction-limited imaging can be achieved with a large aperture (~1.5 m) telescope. Such a telescope has a resolution of < 0.1" in the visible band, and the large aperture means more sensitivity with which to probe the faint universe. Appropriate design of the telescope optics can give diffraction-limited images over a wide field-of-view, which is ideal for survey experiments.

The science case for space-like imaging from an airship is very broad: the majority of the time on the Hubble Space Telescope is allocated to imaging, which demonstrates a high demand for space-like imaging performance in the broader astronomy community. Examples of projects that benefit greatly by using a 1-2 m class optical telescope in near-space conditions (i.e., on an airship platform) include:

- Seeking the first galaxies, stars, and primordial structure during the era of re-ionization with high-redshifted emission which is blocked by the atmosphere for ground-based observatories,

- Imaging of the low surface-brightness sky, wide-field surveying for outstanding problems in "near-field cosmology," e.g. finding Milky Way's satellites,

- Wide-field imaging for dark sector cosmology, including the mapping of clusters, baryonic acoustic oscillations and weak lensing.

An experiment that derives particular benefit from large-scale, high resolution mapping of the sky is weak lensing of large-scale structure and of galaxy clusters. Weak lensing requires a high image quality on sub-arcsecond scales in order to perform accurate galaxy shape measurements. By surveying large areas we can derive weak lensing catalogs that can be used to map the distribution of dark matter over large scales and in galaxy clusters with high fidelity. This will help us to understand what dark matter is, how it is distributed, and how it has influenced the evolution of our universe. Similarly, we will be able to measure galaxy morphology for a vast number of galaxies and their morphological evolution.

**Requirements**

The greatest obstacle to achieving diffraction-limited imaging on an airborne platform is pointing stability. Fortunately, the problem of pointing stability is shared by the balloon community, and projects such as STABLE (Sub-arcsecond Telescope And BaLloon Experiment) at JPL will demonstrate technology that achieves a pointing stability of better than 0.1". It will also be necessary to position the telescope on the airship in such a way that the imaging quality is not affected by the airship, i.e. that the turbulence caused by heat radiating from the airship skin does not degrade the seeing.

**Pathfinder Mission**

A pathfinder mission would carry a 1 m diameter telescope observing in the visible band with a moderate field of view. It will be equipped with an imaging camera that can be used to map an area of the sky with high resolution and deep sensitivity. This pathfinder mission would demonstrate two key interconnected factors:

- It is possible to perform adequate pointing of the telescope to achieve diffraction-limited imaging.
- The observing conditions are indeed stable enough to achieve diffraction-limited imaging for a scientifically useful amount of time.

**Large Scale Option**

When the Hubble Space Telescope's hardware inevitably fails it will not be possible to repair it, and there will be no large visible-wavelength space observatory accessible by the general community. The high demand for Hubble imaging time shows that this will mark a tremendous loss for the field.

With that in mind, the long-term goal is to have a permanent airship observatory hosting a 2-3 m-class telescope (or as large as the airship payload capacity allows) that operates continuously with a broad suite of instruments. For full-sky coverage it will need to be located on the Equator, or there will need to be an airship observatory in each hemisphere. This will be an open facility on which the astronomy community can propose for time. The fact that it is hosted on an airship means that instrument repairs and upgrades can be performed with relative ease compared to space.

# Chapter 5:
# The Path Forward

## Consensus Recommendations

The participants of this study were able to reach a consensus on three major areas (I) for science aboard stratospheric airships, (II) for science aboard low-to-mid altitude airships, and (III) an unexpected splinter topic regarding stratospheric tethered aerostats, recommended for immediate follow-on:

I.  **A. Establish a roadmap toward >60 kft observatory platforms** for Earth, atmospheric and space sciences. We found these platforms to be highly desirable and well-motivated. To make progress, we envision a roadmap as follows:

    i. **Demonstrate** high-altitude airships as a viable platform solution to capability gaps via a prize/challenge.

    ii. **Launch path-finder(s)** for science including site survey and new stratospheric instrument technology.

    iii. **Develop and launch high-altitude, stratospheric observatory(ies).**

    **B. Build a consortium** to educate the wider scientific community about the scientific potential of affordable stratospheric platforms and to further communicate to industry the needs of scientists. To expand what has already begun with this study, we recommend:

    i. **Holding dedicated sessions at American Institute of Aeronautics and Astronautics (AIAA) conferences** where the scientific community presents its relatively flexible needs/requirements (as compared to the military) to the lighter-than-air industry.

    ii. **Sponsor sessions at American Astronomical Society (AAS) and American Geophysical Union (AGU) conferences** on airship capabilities for scientists in the space and Earth science communities.

II. **A. Identify and develop existing airships as science platforms immediately** to be leveraged for the well-motivated Earth and Atmospheric science outlined in earlier chapters.

    **B. Consortium-build to move low-altitude airships to mid-altitudes** for improved capability-gap solutions in Earth and atmospheric science observations.

III. **Develop the first successful stratospheric tethered aerostat platform** to support many of the high-altitude airship platform science goals at potentially an order-of-magnitude less cost (Figure 5.1).

*Airships: A New Horizon for Science*                                                                                                   43

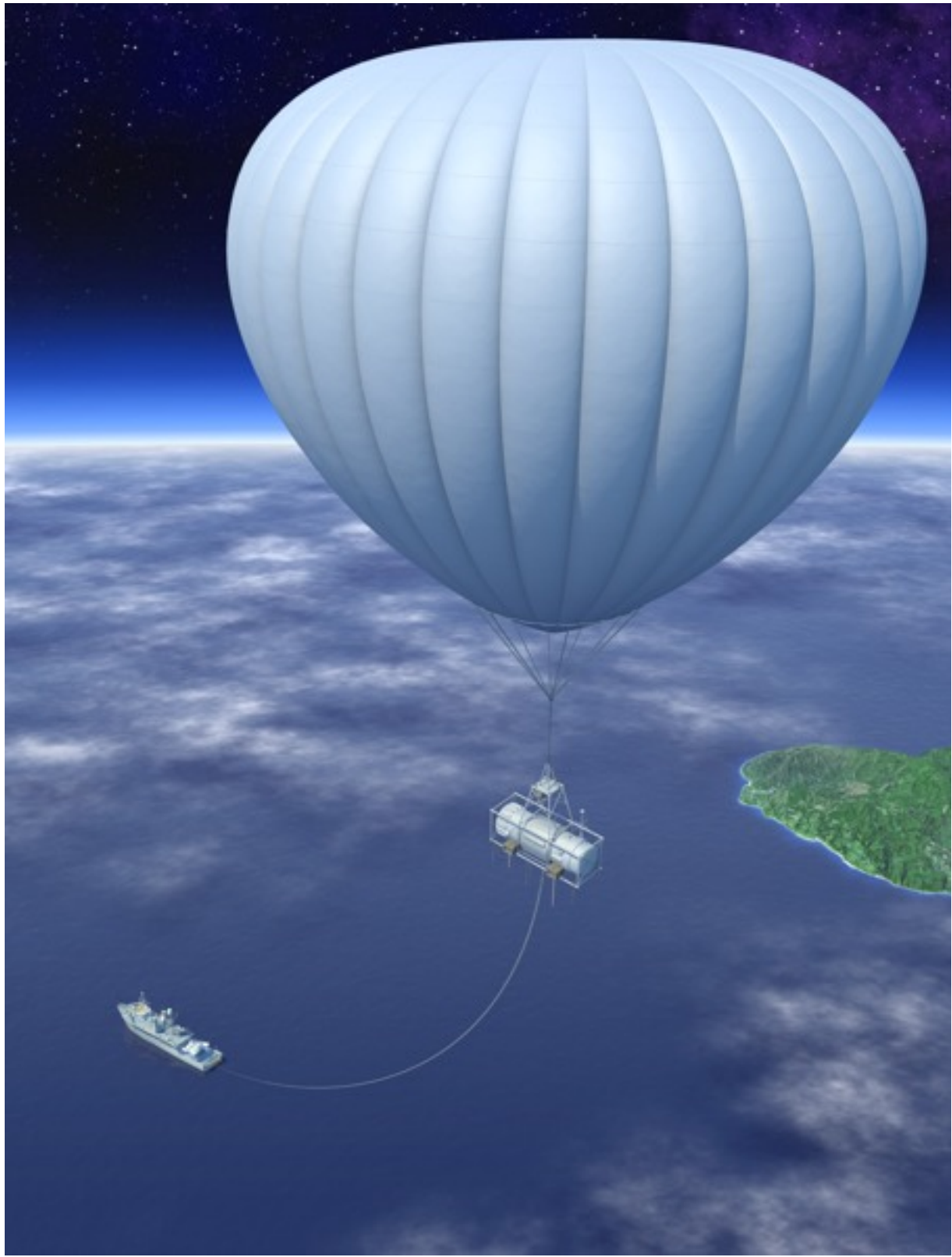

***Figure 5.1***: *An illustrated concept of a stratospheric tethered aerostat (anchored by sea/ship) with a scientific payload. Mike Hughes (Eagre Interactive) / Keck Institute for Space Studies*

# I. A Roadmap to Affordable Airship Platforms in the Stratosphere

The preceding materials and cases in this report have demonstrated significant interest in high-altitude airships as a science platform from the academic community and possible industrial partners. After bringing together NASA, Universities and the private sector (including both large defense contractors and smaller specialized firms) to discuss the capabilities and requirements for airships as a science platform, our consensus is that airships, especially stratospheric/high-altitude airships, open up an exciting new platform for both Earth and space science. Much of the needed technology has been developed via DoD funding over the past decade. However, a persistent (~1 day) airship flight at stratospheric (> 65 kft) altitudes has not yet been demonstrated since DoD funding for airships largely ceased when troop withdrawal from the Middle East became a reality. Among the industrial participants (which included representatives from Lockheed Martin, Northrop Grumman, Aeros, Southwest Research Institute, AeroVironment, and Near Space Corporation) there was **unanimous** consensus that challenge with a substantial prize would spur the investment needed to demonstrate sustained stratospheric airship capabilities.

**The Prize-driven Challenge:** We recommend a million-dollar-class prize for the first organization to fly a powered airship that remains stationary at 65 kft (**20 km**) for over **20 hours** with a **20 kg** science payload (20-20-20; alternatively this could be pushed to 24-24-24 after further consideration in the challenge design phase). The design must be affordably scalable to longer flights with more massive payloads in order to spur further investment by interested agencies and industries.

## Motivations for Producing a Scalable Stratospheric Airship Via a Challenge Scenario

### For Earth and Atmospheric Science

Introduced in Chapter 1 and exemplified in Chapter 4, there are currently many outstanding questions in Earth science which require long duration observations from the stratosphere, beyond what is possible with balloons, aircraft, and satellites. These include:

- Intercontinental transport of pollution
- Exchange of water vapor and ozone-related gases across the tropopause
- Climate feedbacks on stratospheric dynamics
- Evolution of springtime polar ozone loss

### For Space Sciences

Similarly, the list of astronomical applications for an airship is limited only by the ingenuity and vision of the community at large, from broad surveys in unexplored wavelength space to compelling new instrumental designs, including:

- Hubble-competitive imaging of the low surface-brightness sky, wide-field surveys for outstanding problems in "near-field cosmology," finding Milky Way's satellites; dark sector cosmology with large weak lensing surveys.
- Accessing spectral regions unseen by ALMA and beyond the reach of SOFIA, including the exploration and discovery of the Terahertz sky.

- Seeking out primordial structure and re-ionization epoch star formation with extremely high-redshifted emission which is blocked by the atmosphere for ground-based observatories.

- Interferometric imaging of protoplanetary disks, the event horizons of supermassive black holes and other compact objects in our galaxy as well as our galactic neighbors.

- More flexible technology development in space-like conditions without space-like difficulty for new detector technology, probing, e.g., the epoch of inflation.

**Commercial Interest**

In addition to science applications, long duration stratospheric airships would have non-science applications, and various companies have expressed interest in the commercial availability of airships with the capabilities that this challenge would demonstrate. As part of our study we reached out to companies in telecommunications who have major holes in their signal coverage for remote areas and emergency response, as well as the oil and gas industry, who are interested in airship technology for alarm monitoring, asset tracking, field communications and production automation. Additionally, transport companies pay high prices for satellite coverage to aid in filling an incomplete monitoring network across more remote regions. The regional coverage of a stratospheric airship would provide an affordable communications service without holes for tracking cargo in even the most rural areas, which is currently unavailable or cost prohibitive. The industry needs are clear across many sectors, including utilities companies, field communications and locating services, as well as forestry and border control services.

**Defense Applications**

While spending for the theaters of war in Iraq and Afghanistan has subsided considerably, leaving the development of a new generation of stratospheric airships on the brink of realization, the long-term need for these platforms for defense purposes is not gone. The Operationally Responsive Space (ORS) Office is tasked with organizing time-critical response capability if national space capabilities need augmentation or reconstitution. If the TRL of the airship technology was raised through non-military applications, the ORS organizational structure is willing to invest in airship technology as a replacement capability for communications and tracking in this "emergency room" scenario for space.

**Broad and Diverse Interest in Demonstrated Stratospheric Airship Technology**

Between the interest from defense operationally responsive systems, academic and national lab science, and representatives from private industry, we have experienced a wide and open array of interest to contributing funds towards a challenge to deliver a product of commercial and scientific use. These potential partners recognize the motivational power of a prize competition spurring further development. We have also begun seeking out private donors and believe some combination of corporate and private funding could provide a significant portion of the prize purse or operational expenses for this challenge. Clearly the public imagination is excited by the futuristic dream of high-altitude airships, providing the tranquil and sustainable antidote to the traditional fixed-wing aircraft that require many times the power/fuel for payload support. Pushing this technology with the needs of science and space exploration, rather than for war, has clear appeal to the public, as well as the private foundations which could provide matching funds for the stated goals.

## A Centennial Challenge Opportunity at NASA

In NASA's constrained budget environment there are few opportunities for space missions in astronomy and Earth science and these have very long lead times. Traditionally, space data (and near space-like data) have also been acquired via NASA's suborbital program, which includes sounding rockets and balloons. However, airships (powered, maneuverable, lighter-than-air vehicles) could offer significant gains in observing time, sky and ground coverage, data downlink capability, and continuity of observations over existing suborbital options at competitive prices. We seek to spur private industry (or non-profit institutions, including FFRDCs and Universities) to demonstrate the capability for sustained airship flights for astronomy and Earth science.

**NASA Technology Roadmap.** Succeeding in the 20-20-20 Airships Challenge will likely require technical progress in areas that are called out in NASA's space technology roadmap. Some examples include:

- **TA12 :** Materials, Structures, Mechanical Systems and Manufacturing. Lightweight, strong, innovative materials and manufacturing processes for those materials are needed. Likewise, progress in generating solar power is needed for long duration airship flight and spaceflight.

- **TA14:** Thermal Management Systems. The diurnal cycle is a technological challenge for airships and technology developed for thermal management in the near-vacuum of the stratosphere could be brought to bear on space systems.

Likewise, a successful airship challenge would provide a platform on which testing and development of numerous space technology systems would be enabled:

- **TA05 :** Communication and Navigation Systems. Navigation systems, sensors, transmitters, and transponders will need to be tested over long distances and through the atmosphere without the inflexibility of sending them into space and not being able to recover them. Likewise, better atmospheric models are needed to further refine the proposed systems. Airships can gather atmospheric data both in-situ and via remote sensing.

- **TA08:** Science Instruments, Observatories and Autonomous systems. Testing instruments, detectors, deployable mirrors and other technology in a space-like environment would be enabled by an airship.

In coordination with the FAA, NASA would be a natural candidate to run a stratospheric airship challenge for these reasons:

1. There is significant, National Academy-endorsed Earth and space science that is made accessible and affordable by a stratospheric Airship.

2. There is clear commercial and DoD interest in developing a stratospheric airship, that would be spurred by a prize challenge.

3. Private industry has an interest in utilizing the resulting product and is willing to support a challenge financially.

4. A stratospheric airship challenge provides a venue for progress on some NASA Space Technology Goals and a path toward a testing platform for multiple technologies in the future.

## What would a Challenge Competition look like?

Ongoing discussions with industry representatives, who have been part of the diverse participant list of our workshops and study, have clearly reasoned why a prize at less than $1 million USD would not be enough for both new and established teams to take on the risk of a challenge. However most industry representatives considered a prize between $1 and 2 million dollars to be realistic for the design work, materials, labor, and trials needed to be successful. A minority of representatives believed that a prize for over $2 million USD would be required; however that the prize could be less if the winners had explicit, sizeable contracts waiting from industry stake-holders (e.g., large telecommunications or energy companies). All consultants agreed that the prize would not need to exceed $5 million USD, but that it would likely spur at least 10 participants at the $1-2 million USD level.

Technologically, tall poles have included the handling of the diurnal cycle (hence the 20 hour requirement), as well as the sustained payload support at altitude (at least 20 hour duration with at least 20 kg). The 20 km level is set by the stratospheric wind minimum which occurs at that altitude, and together with being above >95% of the atmosphere, suggests the ideal height for sustained, long-term science and telecom platforms. No airship has maintained flight at this altitude for more than 7 hours in the few attempts that have previously occurred. The technological challenges that exist for airships in the stratosphere are considerably different to traditional, low-altitude airship technology, making a stratospheric airship a completely different vehicle to a low-altitude blimp. Industry developers of this technology have simply not had enough opportunities to see this technology through on the one-shot contract model, and those who have been closest to the targeted goals have been underfunded. Military support for previous attempts have been well-funded, but for only a handful of attempts which included extremely narrow requirements on incredibly short timelines (that were doomed to fail and did). Allowing teams to have more than two years of development and trial time for the much less stringent operational requirements will very likely lead to success in 3-5 years, although there are a few teams which could potentially compete and win within two years of a challenge launch.

**Work Plan.** Principal investigator(s) and project manager(s) would direct the work of the team over a 6 month period to define the rules and timeline for the competition, continue ongoing efforts to generate interest in the community of potential competitors, formalize efforts to raise money for a challenge, reaching out to the wider astrophysics and Earth science community to generate interest and excitement in both the challenge and the resulting airship platform.

Much of the challenge definition work is straightforward and will flow naturally out of efforts and relationships formed via this study. However, two key aspects of the challenge must be resolved in this 6 month period. The first are the rules defining scalability. The ultimate goal is to create a platform that can achieve sustained flight for weeks or months at a time and can carry a payload of hundreds or thousands of kilograms. The rules of such a challenge must be flexible enough to allow multiple innovative approaches, but specific enough to require that a winning entry has a path towards scalability in duration and payload capacity.

The second key aspect of the challenge definition is to scope out the 20 kg payload, define how this will be constructed and what the interface between the payload and the vehicle will be. This could be a small science instrument capable of making rudimentary in-situ science observations of stratospheric atmospheric conditions and in taking turbulence and wind shear measurements during ascent and descent. Recovery of this instrument and the resulting data will be a condition of success in the challenge. A full design for the payload must be identified, along with mechanical, electrical, and thermal interfaces to the potential airship platforms delineated. This includes plans for construction of one or more science payloads to be used during the competition. A related task will be to define how the challenge will be monitored and judged. Since an externally provided payload will be needed for a competitor to complete the challenge, procedures for interaction between the challenge organizers and competitors must be defined. This will occur in conjunction with potential competitors with whom a dialog has already begun.

Other rules and regulations to be defined during the 6 month period are:

- Length of challenge (we envision 2-5 years).
- Requirements on platform stability and "loiter area" (i.e. the radius the Airship is allowed to "wander" within).
- An appropriate FAA sanctioned "no fly zone" for airship tests and challenge attempts. One potential defense contractor contestant already has such an area and has offered to open this up to other potential challenge attempts.

Expertise of JPL's suborbital community, particularly its ballooning experts from the Earth science, planetary science and astrophysics communities will continue to be sought in the further development of this challenge. A potentially relevant JPL project being developed right now called STABLE (Sub-arcsecond Telescope And BalLoon Experiment), can provide local expertise to help evaluate challenge participants.

A proposal along these lines has already been submitted to NASA by a sub-team (PI: Rhodes) of this study, for the further development of the stratospheric airship challenge.

## Open Questions for Specific "Stratospheric Observatory" Concepts Long-term

### What will "an observatory" aboard an airship weigh?

This open question motivates the development of lightweight instrumentation without the extreme restrictions of nano-satellites and/or rocket launch transport.

### What will it cost?

If a sizable market across defense and communications industries can be spurred by a successful challenge competition participation, then costs will be driven to realistic levels for a competitive platform in the science budgetary environment.

### Should space, Earth, and atmospheric sciences all share the same platform?

A shared platform between vastly different observational modes for different science goals will be significantly more complicated, however specific harmonious requirements between Earth-ward and Space-ward science would be possible to design on a concept-by-concept basis. Ultimately cost may drive the necessity for a multi-disciplinary platform, as well as the specific funding body for a given airship observatory concept.

**Sustaining a Community Base for Continued Development**

Given the success of this study to gather leaders across the relevant fields to examine the opportunity of science aboard airships, it will be important moving forward to communicate the findings of this study and grow the community that has been seeded by this study. The workshops and meetings which took place at the Keck Institute for Space Studies over the study period not only strengthened the bonds between members of NASA JPL, the Caltech campus, external institutions and industry, but also strengthened interactions between traditionally separate scientific disciplines, which could see unprecedented levels of collaboration aboard multi-disciplinary airship platforms. Nurturing this budding community will be greatly aided by the building of a consortium which aims to further inform and identify airship science opportunities. Recommendations include:

- **Holding dedicated sessions at AIAA conferences** where the scientific community presents its relatively flexible needs/requirements (as compared to the military) to the lighter-than-air industry.
- **Sponsor AAS and AGU sessions** on airship capabilities for scientists in the space and Earth science communities.

## II. Utilization of Existing Low-to-Mid Altitude Airships

A new community of scientists interested in low-to-mid altitude airship platforms has already been seeded by this study. While this is not a recommended area of Keck Institute for Space Studies technical development follow-on, given that these technologies are already at an advanced stage, the study recognizes that current airship platforms are under-utilized given their unique capabilities in Earth and atmospheric science. Because of their persistence, these airships fill a scale gap in between "anecdotal" ground-based or aircraft measurements and global measurements from satellites.

Low-altitude (< 20 kft) piloted Airships have long been recognized as excellent platforms for Earth observations. For example, the European Union-funded research project PEGASOS (Pan-European Gas AeroSOl Climate Interaction Study) flew an airship for 20 weeks in 2012 across parts of Europe analyzing air chemistry. More recently, the British Broadcasting Corporation's CloudLab Airship flew in September and October of 2013 from Orlando, Florida to San Francisco, California investigating aerosols and clouds.

For Earth and atmospheric science, our consensus is that the only measurements that *require* a high-altitude rather than medium-altitude airship are those focused on the upper troposphere / lower stratosphere or on extreme weather (this is of course not the case with astronomy and astrophysics). For some of the most compelling science cases (i.e., megacities, ecosystems, and coastal ocean studies) high-altitude airships provide a "bigger picture" and longer timescales, but they also allow less payload for doing multidisciplinary studies. The medium-altitude airships (20-40 kft) do not provide the same large-scale picture (at least not without a flying pattern) and may require revisits for longer temporal coverage, however projected capabilities enable large and flexible instrument payloads.

Growing a consortium to more precisely identify and utilize existing airship platforms is recommended by this study and should be expedited and supported across a variety of means, both public and private. Existing airships which can support the unique needs for various airborne science programs include:

- Expanding the NASA fleet to include the US Navy Airship **MZ-3**
- Northrop Grumman **M1400** (with potential for operating at 20-40 kft)

## III. Summary and Conclusions Regarding Stratospheric Tethered Aerostats

During the course of the workshop, it became clear to the assembled study team that the alternative technology of tethered aerostats was a possible, but unproven, option for meeting many of the science objectives of interest to the team. The key realization was that the very difficult challenge of power and propulsion for free flying airships could be avoided if the vehicle were anchored to the ground via a long tether. Such an approach necessarily applies only to those applications that are satisfied with a stationary platform.

A subset of the overall study team (Fesen, Goldsmith, Hall, Lachenmeier, Lord, Miller, Rhodes, and S. Smith) pursued the tethered aerostat option after the opening workshop through teleconferences, literature searches, back of the envelope calculations and ongoing discussions. It became clear that tethered aerostats have their own substantial technical challenges centered on the related problems of initial deployment, tolerance to winds and weather, minimization of tether size and mass, and relatively low payload to overall system mass ratios. These issues and others were subsequently discussed at a supplementary one-day meeting hosted by KISS in mid-November (2013) that specifically focused on the tethered aerostat option. Three additional experts joined that meeting: Mike Smith from Raven Industries, Gil Baird from ILC Dover, and Sara Smoot, a recent graduate of Stanford University who did a PhD thesis involving high altitude tethered vehicles (theoretical and experimental).

The basic conclusion of the one-day supplementary meeting is that there is insufficient information to know if a stratospheric tethered aerostat can be made to work for long durations or for a cost that is attractive to future users. Many paper studies have been done over the years, but no flight tests have been attempted since some partially successful experiments by the French in the early 1970s. A key reason for the 40 year lack of flight experiments is that the smallest possible aerostat must still be rather large (~ 5000 m$^3$) to lift itself and the tether to the desired 65+ kft float altitude. Such a large system is expensive to fabricate and test, forming a barrier to continued experimental development. All meeting participants were in agreement that, however expensive, it is of vital importance to perform this kind of stratospheric flight experiment in order to properly evaluate the feasibility of the stratospheric tethered aerostat concept.

Another noteworthy outcome of the supplementary one-day meeting is that there is not just one tethered concept but instead a spectrum of possible variations. These include:

- The conventional single stratospheric balloon tethered to the ground.

- Multi-balloon architectures that use lower altitude balloons to carry some of the tether weight and thereby reduce the size and mass of the stratospheric platform and potentially of the overall system.

- Tethering the balloon(s) to a ship at sea both as an aid to initial deployment, where the ship can move with the winds and help reduce the aerodynamic drag loads, and to provide mobility for the system to avoid bad weather or to observe from a different location.

- In contrast to the conventional deployment approach of having the tether always connected to the ground and the aerostat, use the alternative of having the aerostat drop the tether to the ground once at altitude. This requires a separate vehicle (ship, truck) to collect the tether once it reaches the ground and anchor it.

- Tethered multi-balloon concepts that are not connected to the ground but are connected to each other. This concept relies on drag modulation capability on each balloon and different winds at different altitudes to provide station-keeping without the need for propulsion.

- Tethering a high-altitude, stratospheric airship or balloon to a lower altitude, tropospheric "tug" vehicle (Fesen and Brown, in prep.) making use of the east/west wind shear between the stratosphere and upper troposphere to keep the upper platform on station and carrying a science payload (see Figure 5.2).

- Adding lift modulation (wings) to any of the above concepts to help carry tether weight via aerodynamic lift instead of buoyancy.

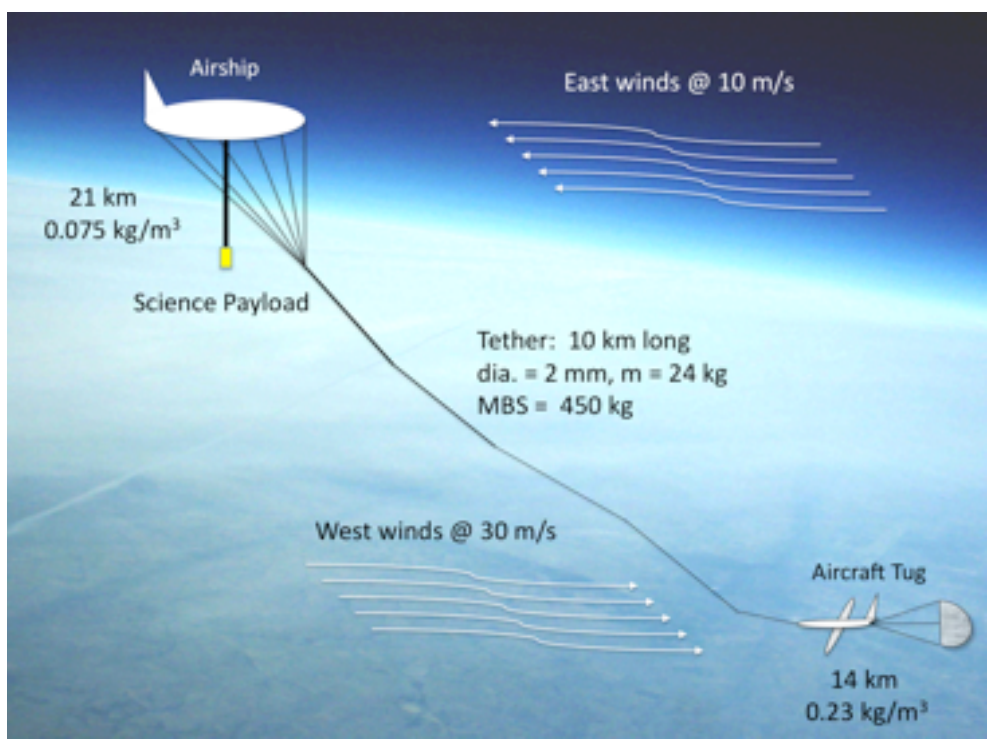

*Figure 5.2*: Cartoon sketch of the basic idea behind a stratospheric, station-keeping LTA science platform using East--West wind shear between the stratosphere and troposphere. Listed values are only meant as representative numbers. Station-keeping of the stratospheric science carrying airship is accomplished through active control of the lower vehicle's drag force by varying it aerodynamic shape and hence its effective form drag coefficient to generate an equal but opposite drag force to than experienced by the upper platform. Some control of both upper and lower vehicles motion perpendicular to wind directions could be done on both vehicles. (Fesen and Brown, in prep.)

## Tethered Aerostat vs. Airship Platform Capabilities

The Earth and atmospheric science component of our study concluded that the main drawback to a tethered platform is that it would greatly limit the area over which you can observe (and prohibit the parameter space of phenomena-following), which is a major advantage to the new capability space airships open up for Earth and atmospheric science. However since most of the sky regions of interest would be visible from a well-selected site, the phenomena-following maneuverability of the platform (beyond station keeping and weather avoidance) is not a key requirement for the overwhelming majority of space science needs. Station-keeping with a tether as opposed to propulsion could be an order of magnitude less expensive platform for most areas of space science and some areas of Earth and atmospheric science. The development of airships remains a unique opportunity for new event and phenomena-following observations, assuming the success of a new stratospheric tethered aerostat platform.

**Recommendations for Stratospheric Tether Development and Future Work Plan**

The tethers sub-team of our study makes the following specific recommendations for future work on the stratospheric tethered aerostat concept:

1) **Analysis, simulation and design** trade work should be done to quantify the expected performance of most or all of the different versions of the tethered aerostat concept list above. The important performance metrics are: system mass, payload mass, operational altitude, wind tolerance, platform attitude stability and operational lifetime.

2) **Perform a risk analysis** on the leading architecture options using the data from #1 above and other expert opinions gathered from the engineering community.

3) **Plan and execute a prototyping and test plan** focused on addressing the top risks identified in #2 above. Budget constraints will dictate how many risks can be addressed, but it should be sufficient to answer the question of which approaches, if any, can meet the technical and cost metrics required to be a valuable scientific observation platform for the space and earth science communities. The test plan should include flying a tethered aerostat at the desired operational altitude.

4) **A science advisory board** should work to develop and refine the science applications to be addressed by the tethered aerostat and generate performance specifications that the overall tethered aerostat system must satisfy, in coordination with the engineering development activity described in #1 to #3 above.

# Final Remarks

Whether we eventually see full-scale stratospheric airships host observatory-class facilities, or more tailored experiments for game-changing science (Recommendations I.A.ii and I.A.iii), we anticipate and aim to nurture the further development of this promising platform (Rec I.B.i and I.B.ii). An industry- and market-spurring, prize-driven challenge will be an excellent place to start (Rec I.A.i). As inspiration for the further development of their high-altitude counterparts, and more presently representing an under-utilized platform requiring no significant technical development for science, we highly recommend leveraging the existing low-altitude airships for remarkably unique Earth and atmospheric science (Rec II.A and II.B). Finally, we recommend for immediate follow-on and technical development, the detailed modeling, designing and building of the first successful stratospheric tethered aerostat platform for the use of compelling Earth, atmospheric, and space sciences (Rec III).

# Acknowledgements

| Report Section | Primary Content Authors and Editors |
|---|---|
| *Executive Summary* | Robert Fesen, Sarah Miller |
| *Chapter 1* | Sarah Miller, Randall Friedl, Robert Fesen, Gregory Quetin, Jeffery Hall, Riley Duren |
| *Chapter 2* | Steve Lord, Dave Carlile, Steve Smith |
| *Chapter 3* | Oliver King, Jessica Neu, Steve Smith, Eliot Young |
| *Chapter 4* | Randall Friedl, Jessica Neu, Oliver King, Geoff Blake, Jeff Booth, Steve Lord, Riley Duren, Marc Simard, Stan Sander, Jason Rhodes, Eliot Young |
| *Chapter 5* | Sarah Miller, Jason Rhodes, Jeffery Hall, Riley Duren, Jessica Neu, Robert Fesen, Brent Freeze |

**Draft comments by**: Jessica Neu, Dave Carlile, Jason Rhodes, Lynne Hillenbrand, Steve Smith, Jeff Hall, Greg Quentin, Steve Lord, Abigail Swann, Tim Lachenmeier, Mike Smith, and Paul Goldsmith

**Meta-editing**: Sarah Miller (and Oliver King on *Chapter 3*), **Document formatting:** Sarah Miller

**Illustrations**: (on p.2 and p.44) Mike Hughes and Chuck Carter from Eagre Interactive (Keck Institute for Space Studies)

While we acknowledge specific contributions to this report above, all study members listed on the title page participated in invaluable ways to our brain-storming discussions, conclusions, and recommendations over the course of the study period. In addition to the study members who contributed to the Short Course presentations (Robert Fesen, Jens Kauffmann, Steve Lord, Randy Friedl, Geoff Blake, Paul Goldsmith, and Sarah Miller), we also thank Michael Werner for his presentation.

The members of this study would like to extend their gratitude to Michele Judd, Managing Director of the Keck Institute for Space Studies, and her excellent team, for creating an optimal working environment during the workshops and meetings of this study. Judd played a pivotal role in the creation of our new airship science community over the course of the study period.

We also thank the director of the Keck Institute for Space Studies, Tom Prince, as well as the Steering Committee for selecting our proposal from the 2013 study program candidates. The Institute provided a unique opportunity to bring key science and industry leaders together to make this comprehensive evaluation and report possible.

We thank the larger science community for their participation during the Short Course of this study, and their continued interest and support in new airship and stratospheric tether platforms for science. We thank the public for their interest and participation in the public component of this study.

We acknowledge NASA's Jet Propulsion Laboratory and the California Institute of Technology for their internal support of the process and program of the Keck Institute for Space Studies.

Finally, we would like to thank the W. M. Keck Foundation for establishing and sustaining the Keck Institute for Space Studies.